\documentclass[floats,floatfix,showpacs,preprintnumbers,amssymb,prd,superscriptaddress,nofootinbib,nolongbibliography,reprint]{revtex4-1}
\usepackage[dvipdfmx]{graphicx}

\usepackage{subfigure}
\usepackage{bm}
\usepackage{mathrsfs}
\usepackage{amsmath}
\usepackage{cases}

\def\be{\begin{equation}}
\def\ee{\end{equation}}
\newcommand{\beq}{\begin{eqnarray}}
\newcommand{\eeq}{\end{eqnarray}} 
\usepackage{color}
\usepackage[normalem]{ulem}
\usepackage[linktocpage=true]{hyperref}
\usepackage{cancel}

\begin{document}
\title{Out-of-Equilibrium Effects in Non-Radial Relativistic Stellar Perturbations: A Model-Agnostic Formulation and Mode Analysis}

\author{Takuya Katagiri}
\email[Corresponding author: ]{takuya.katagiri@uniroma1.it}
\affiliation{Dipartimento di Fisica, Sapienza Università di Roma, Piazzale Aldo Moro 5, 00185, Roma, Italy}
\affiliation{INFN, Sezione di Roma, Piazzale Aldo Moro 2, 00185, Roma, Italy}

\author{Victor Guedes}
\affiliation{Department of Physics, University of Virginia, Charlottesville, Virginia 22904, USA}

\author{Kent Yagi}
\affiliation{Department of Physics, University of Virginia, Charlottesville, Virginia 22904, USA}

\date{\today}

\begin{abstract}
We present a systematic, model-agnostic analysis of out-of-equilibrium effects, including viscosity and thermal conductivity, in non-radial oscillations of relativistic stars. Extending the Lindblom-Detweiler formalism, we construct, to our knowledge, the first general framework for linear, non-radial relativistic stellar perturbations that incorporates generic nonequilibrium corrections to the perfect-fluid sector in both the even- and odd-parity channels. Our framework is formulated in terms of the tensorial structure and thermodynamic decomposition of generic corrections without relying on any specific constitutive relations, thereby allowing us to elucidate, at a structural level, how these effects enter the perturbation equations and contribute to geometric deformations and fluid fluctuations. As an application, we consider the Bemfica-Disconzi-Noronha-Kovtun fluid and perturbatively investigate shifts in the frequencies and damping times of modes connected to their perfect-fluid counterparts in the limit of vanishing transport coefficients. We also identify structural features of the closed eigenvalue problem that suggest the possibility of additional mode families. Our formalism provides a unified framework for analyzing how different relativistic fluid theories modify the structure of non-radial stellar perturbations.

\end{abstract}

\maketitle
\tableofcontents

\section{Introduction}
Gravitational-wave~(GW) astronomy has opened a new observational window onto physics ranging from cosmology to matter~\cite{Somiya:2011np,VIRGO:2014yos,LIGOScientific:2014pky,2017arXiv170200786A,Kasen:2017sxr,LIGOScientific:2017ync,LIGOScientific:2017adf,Romero-Shaw:2020thy,LIGOScientific:2025ttj}. Neutron stars~(NSs) are one of the most promising GW sources and provide an unprecedented opportunity to probe extreme matter at supranuclear densities through GW observations~\cite{Lattimer:2015nhk,Ozel:2016oaf,Burgio:2021vgk}. The composition and dynamics of matter in the interior of NSs, coupled to gravity, involve a wide range of modern physics, such as gravitation, hydrodynamics, superfluidity and superconductivity, electromagnetism, and nuclear physics. Precise theoretical modeling of NSs, therefore, offers a crucial link between fundamental physics and observational signatures.

One of the most significant achievements to date in connecting theory to observations is the first detection of GWs from a binary NS merger,~GW170817, which has placed constraints on the dense matter equation of state and stellar radii through measurements of tidal deformability~\cite{LIGOScientific:2017vwq,LIGOScientific:2018cki,LIGOScientific:2018hze,LIGOScientific:2020aai}. With the rapid development of future GW detectors~\cite{Punturo:2010zz,Isoyama:2018rjb,Reitze:2019iox,Kawamura:2020pcg,ET:2025xjr} and data-analysis frameworks~\cite{Meacher:2015rex,Ashton:2018jfp,Hoy:2020vys,Williams:2022pgn,Tania:2025bsa}, it is reasonable to anticipate precise measurements of NS observables across multiple frequency bands and even through stellar oscillation modes, potentially revealing ``unknown unknowns'' as well. Hence, developing theoretical frameworks of fundamental interactions inside perturbed NSs, incorporating relevant physical processes, is becoming increasingly important for interpreting future observations in terms of the underlying microphysics. 

The theory of stellar perturbations provides a fundamental and powerful framework for investigating the features of GWs emitted by non-radially deformed stars, as well as tidal interactions with nearby external gravitational sources. Information about the internal structure of a star is encoded in the characteristic frequencies and damping rates of the emitted GWs~\cite{Kokkotas:1999bd,Ferrari:2011rb}, thereby giving rise to the field of GW asteroseismology~\cite{Andersson:1997rn,Benhar:2004xg,Andersson:2021qdq}, which has become a central topic in modern astrophysics. A robust relativistic treatment of stellar perturbations was initiated by Thorne and Campolattaro~\cite{1967ApJ...149..591T} and subsequently developed by a series of works~\cite{1970ApJ...159..847C,1969ApJ...158....1T,1969ApJ...158..997T,1973ApJ...181..181I}. For non-rotating perfect-fluid stars, the standard formalism was later established by Lindblom and Detweiler~\cite{Lindblom:1983ps,Detweiler:1985zz}. Several extensions beyond the standard single-perfect-fluid description of non-radial perturbations have been proposed, including formulations incorporating heat transfer and chemical diffusion~\cite{Gualtieri:2004av}, pressure anisotropy~\cite{Doneva:2012rd,Mondal:2023wwo,Lau:2024oik}, more general dissipative fluids~\cite{Redondo-Yuste:2024vdb}, and gravitationally coupled multi fluids~\cite{Kumar:2026tgi}. 

The perfect-fluid description assumed in the standard framework for stellar perturbations is generally believed to provide a good leading-order approximation to the matter dynamics, particularly for cold and old NSs such as those typically found in pre-merger binaries within the GW observational window. Indeed, classical works suggested that stellar internal viscosity has a negligible impact on the GW phase because analyses of tidal angular-momentum exchange showed only small phase corrections~\cite{Lai:1993di}, and non-negligible effects on the orbital evolution require unphysically large viscosity~\cite{Bildsten:1992my}. Recent analyses~\cite{Most:2021zvc,Ripley:2023qxo,Ripley:2023lsq,HegadeKR:2024slr}, however, found that dissipative processes arising from internal fluid viscosity may leave detectable imprints on GW signals due to a large finite-size enhancement of corrections. Furthermore, Ref.~\cite{Ghosh:2023vrx} suggests that internal bulk viscosity involving hyperons could heat the NS core up to $10^9-10^{10}{\rm K}$, potentially leaving measurable imprints on GW signals during the inspiral stage, whereas the impact of weak-interaction-driven bulk viscosity might be too small to be detected even with future GW detectors~\cite{HegadeKR:2026iou}. Several studies~\cite{Alford:2017rxf,Most:2021zvc,Most:2022yhe,Chabanov:2023abq,Chabanov:2023blf} indicate that the bulk viscous dissipation may affect post-merger GW signals from binary NS merger remnants. In addition, neglecting tidal dissipation from hyperon bulk viscosity in waveform modeling may bias parameter estimation for third generation GW detectors~\cite{Ghosh:2025glz}.

Moreover, detailed studies of transport processes in perturbed NSs suggest that the viscous effects may be important in modeling NSs as GW sources. The internal shear and bulk viscosities arise from various microscopic processes and depend sensitively on the local temperature profile inside the star. Shear viscosity is generically associated with momentum transport induced by particle scattering. For typical core temperatures in binary NSs, electron-electron scattering provides the dominant contribution to shear viscosity~\cite{1976ApJ...206..218F,1987ApJ...314..234C}, whereas other processes, such as neutron-neutron and neutron-muon scattering, are subdominant~\cite{Shternin:2008es}. Bulk viscosity, by contrast, can arise from direct and modified Urca reactions in NS interiors and becomes important at high temperature. In the presence of non-leptonic weak processes, tidal heating can heat up a NS to sufficiently high temperature for hyperon bulk viscosity to dominate the shear viscosity due to electron-electron scattering~\cite{Ghosh:2023vrx}.

From the theoretical perspective of relativistic fluid dynamics, the behavior of viscous fluids remains less understood, especially in curved spacetime. The history of relativistic formulations of viscous hydrodynamics has been far from straightforward, and many proposed theories have suffered from drawbacks incompatible with fundamental physical principles: (i)~stability, (ii)~causality, (iii)~local well-posedness, and (iv)~strong hyperbolicity. The earliest formulations were proposed by Eckart~\cite{PhysRev.58.919} and by Landau and Lifshitz~\cite{Landau1987Fluid}; however, these theories predict that a global equilibrium state is unstable and even exhibit superluminal propagation~\cite{PhysRevD.31.725}, thereby violating the conditions~(i) and (ii). Later, M{\"u}ller-Israel-Stewart theory~\cite{1967ZPhy..198..329M,Israel:1976tn,Israel:1976efz,1977RSPSA.357...59S,1979RSPSA.365...43I,Israel:1979wp} resolved the issues of stability and causality by introducing additional dynamical degrees of freedom. Nevertheless, the conditions~(ii)--(iv) have been established only in restricted settings~\cite{Bemfica:2019cop,Brito:2020nou,Cordeiro:2025rkt,Cordeiro:2025mtg}. 

More recently, Bemfica, Disconzi, and Noronha~\cite{Bemfica:2017wps} constructed a first-order theory of relativistic hydrodynamics for conformal fluids, based on a gradient expansion about the perfect-fluid constitutive relations, and rigorously established that the conditions~(i)--(iii) hold. Kovtun~\cite{Kovtun:2019hdm} subsequently demonstrated the existence of linearly stable hydrodynamic frames for nonconformal fluids without a conserved particle current. Hoult and Kovtun~\cite{Hoult:2020eho} then extended this analysis to fluids with a conserved particle current and chemical potentials. Bemfica, Disconzi, and Noronha~\cite{Bemfica:2020zjp} developed a fully general-relativistic formulation incorporating shear and bulk viscosities, thermal conductivity, as well as nonzero baryon density, and rigorously established that it can satisfy the conditions~(i)--(iv) under appropriate constraints on the transport coefficients. This class of first-order theories is commonly referred to as the Bemfica-Disconzi-Noronha-Kovtun~(BDNK) hydrodynamics. For the BDNK fluid, radial stellar oscillations and non-radial axial stellar perturbations have recently been studied~\cite{Caballero:2025omv,Shum:2025jnl,Mendes:2025oib,Keeble:2026bzo,Bussieres:2026rnz}, suggesting nontrivial features of the resulting characteristic oscillations.

Motivated by astrophysical and theoretical hydrodynamical considerations, in this work we perform a systematic, model-agnostic analysis of out-of-equilibrium effects in non-radial perturbations of spherically symmetric relativistic stars. To this end, we construct, to our knowledge, the first general framework for linear, non-radial relativistic stellar perturbations that incorporates generic nonequilibrium corrections to the perfect-fluid sector in both the even- and odd-parity channels, by extending the Lindblom-Detweiler formalism~\cite{Lindblom:1983ps,Detweiler:1985zz}. As a concrete example, we apply our formalism to the BDNK fluid and analyze how the out-of-equilibrium contributions influence non-radial stellar oscillations in terms of both structure of the perturbation equations and the resulting shifts in the characteristic mode frequencies. The BDNK fluid model is particularly convenient for demonstrating our framework because of its relative simplicity to other causal models, such as M{\"u}ller-Israel-Stewart fluids~\cite{1967ZPhy..198..329M,Israel:1976tn,Israel:1976efz,1977RSPSA.357...59S,1979RSPSA.365...43I,Israel:1979wp}, in the sense that it does not require to introduce independent relaxation fields and their associated evolution equations. It is worth noting that our model-agnostic formulation remains applicable to other formulations, including M{\"u}ller-Israel-Stewart fluids. Our formalism provides perturbation equations for generic out-of-equilibrium corrections~(see Eq.~\eqref{eq:fieldeqs} and~\eqref{eq:baryonnumberconservation}), while additional evolution or constraint equations for independent dynamical variables must be supplied separately, when present.

The remainder of this paper is organized as follows. In Section~\ref{Section:PerturbationTheory}, we briefly introduce the overview of our framework. In Section~\ref{Section:EvenParity}, we formulate the even-parity perturbation and discuss the impact of out-of-equilibrium effects on geometric deformations and fluid fluctuations. In Section~\ref{Section:OddParity}, we introduce the formalism for the odd-parity perturbation and elucidate its structure. In Section~\ref{Section:Application}, we apply our formalism to the BDNK fluid and discuss the impact of the out-of-equilibrium contributions on characteristic mode frequencies. Section~\ref{Section:Summary} is devoted to the summary of the present work. In Appendix~\ref{Appendix:Derivation}, we describe the detailed derivation of the even-parity perturbation equations. In Appendix~\ref{Appendix:BDNK}, we provide the mapping to BDNK fluids and outline a practical procedure for solving the perturbation equations. We adopt the geometric units,~$c=G=1$, throughout the paper.

\section{Framework}\label{Section:PerturbationTheory}

\subsection{Setup}
We consider a background, spherically symmetric star in hydrodynamical, thermal, and chemical equilibrium within a perfect-fluid approximation. The line element in Schwarzschild-like coordinates~$(t,r,\theta,\varphi)$ reads
\begin{align}
    g_{\mu \nu} dx^\mu dx^\nu=-e^{\nu}dt^2+e^{\lambda} dr^2+r^2\Omega_{AB}dx^Adx^B,
\end{align}
where $\nu=\nu(r)$, $\lambda=\lambda(r)$, and $\Omega_{AB}dx^Adx^B= d\theta^2+\sin^2\theta d\varphi^2$. The stellar solution satisfies
Einstein's equations,
\begin{align}
    G_{\mu\nu}=8\pi T_{\mu\nu},\label{eq:EinsteinEqscov}
\end{align}
with the Einstein tensor~$G_{\mu\nu}$ and the stress-energy tensor of a perfect fluid,
\begin{align}
    T_{\mu\nu}= \varepsilon u_\mu  u_\nu +p \Delta_{\mu\nu}.
\end{align}
Here, $\varepsilon=\varepsilon(r)$ and $p=p(r)$ are total energy density and pressure, respectively. The vector field~$u^\mu$ is the four-velocity of the fluid, normalized by $g_{\mu\nu}u^\mu u^\nu=-1$, and hence, $u^\mu=(e^{-\nu/2},0,0,0)$. The tensor field~$\Delta_{\mu\nu}:=u_\mu u_\nu +g_{\mu\nu}$ is a projection operator, mapping any tensor field onto hypersurface perpendicular to $u^\mu$. 

Einstein's equations~\eqref{eq:EinsteinEqscov} reduce to a set of the following Tolman-Oppenheimer-Volkoff equations:
\begin{align}
    m'=&4\pi  r^2\varepsilon,~~\nu'=\frac{2m+ 8\pi r^3 p}{r\left(r-2m\right)},\nonumber\\
    p'=&-\left(\varepsilon+p\right)\frac{m+4\pi r^3 p}{r\left(r-2m\right)},~\label{eq:EinsteinEqs}
\end{align}
where the prime denotes a differentiation with respect to $r$. The function~$m:=(r/2)(1-e^{-\lambda})$ represents a radial mass function. The exterior solution to Eq.~\eqref{eq:EinsteinEqs}, where $\varepsilon=p=0$, is the Schwarzschild metric~$e^\nu=e^{-\lambda}=1-{2M}/{r}$, where $M$ is the gravitational mass. The system requires an internal relation between pressure and energy density, i.e., the equation of state. We do not impose any particular equation of state at the current stage.

Now, we consider time-dependent deformation of the star within linear stellar perturbation theory. The perturbed configuration deviates from equilibrium governed by Eq.~\eqref{eq:EinsteinEqs}, inducing out-of-equilibrium effects. To describe the out-of-equilibrium corrections to the perfect fluid part in a model-agnostic manner, we introduce a second-rank symmetric tensor field~$\mathscr{S}_{\mu\nu}$ and a vector field~$\mathscr{J}^\mu$ without specifying their microscopic origin. These tensor and vector fields correspond to the out-of-equilibrium part of the stress-energy tensor and the baryon number current density, respectively. The equations for the perturbed configuration are then given by the linearized Einstein equations and the stress-energy conservation,
\begin{align}
    \delta G_{\mu \nu}=8\pi \left(\delta T_{\mu \nu}+\mathscr{S}_{\mu\nu}\right),\quad \delta \left(\nabla_\mu T^{\mu \nu}\right) =-\nabla_\mu \mathscr{S}^{\mu \nu},\label{eq:fieldeqs}
\end{align}
and the perturbed baryon number conservation,
\begin{align}
    \delta \left[ \nabla_\mu \left(n u^\mu \right)\right]=-\nabla_\mu \mathscr{J}^\mu,\label{eq:baryonnumberconservation}
\end{align}
where $\delta$ denotes the Eulerian displacement of the quantity; $n$ is the baryon number density in the configuration associated with the perfect fluid. The perfect-fluid piece, $T_{\mu\nu}$ and $n u^\mu$, is not conserved alone in the perturbed configuration due to the presence of the inflow to and outflow from a fluid element comoving with $u^\mu$ but the total quantities,~$T_{\mu\nu}+\mathscr{S}_{\mu \nu}$ and~$n u^\mu+\mathscr{J}^\mu$, are conserved to linear order.

The tensor field~$\mathscr{S}_{\mu\nu}$ and the vector field~$\mathscr{J}^\mu$ are, in general, decomposed into
the following form~\cite{Kovtun:2012rj,Kovtun:2019hdm}:
\begin{align}
    \mathscr{S}_{\mu\nu}=&{\cal E} u_\mu u_\nu +{\cal P}\Delta_{\mu\nu}+ {\cal Q}_\mu u_\nu +{\cal Q}_\nu u_\mu +{\cal T}_{\mu\nu},\label{eq:Smunu}\\
    \mathscr{J}^\mu=&{\cal N}u^\mu+{\cal J}^\mu,\label{eq:Jmu}
\end{align}
where ${\cal E}:=u^\mu u^\nu\mathscr{S}_{\mu\nu}$ is the out-of-equilibrium energy density; ${\cal P}:= (1/3)\Delta^{\mu\nu}\mathscr{S}_{\mu\nu}$ is the out-of-equilibrium pressure; ${\cal Q}_\mu:=-\Delta_{\mu}^\rho u^\nu \mathscr{S}_{\rho\nu}$ is heat flux; ${\cal T}_{\mu\nu}:=[\Delta_\mu^\rho \Delta_\nu^\sigma -(1/3)\Delta^{\rho\sigma}\Delta_{\mu\nu}]\mathscr{S_{\rho\sigma}}$ is the shear stress; ${\cal N}=-u_\mu \mathscr{J}^\mu$ is the out-of-equilibrium baryon number density ; and ${\cal J}_\mu=\Delta_{\mu\nu}\mathscr{J}^\nu$ is the out-of-equilibrium diffusion current. A choice of ~$({\cal E},{\cal P},{\cal Q}^\mu,{\cal T}^{\mu\nu},{\cal N},{\cal J}^\mu)$, commonly referred to as a hydrodynamic frame, is determined by the constitutive relations for a given fluid model. The vector fields,~${\cal Q}^\mu$ and ${\cal J}^\mu$, are transverse, i.e., $u_\mu{\cal Q}^\mu=0=u_\mu {\cal J}^\mu$, while the tensor field~${\cal T}_{\mu\nu}$ is transverse, symmetric, and traceless, i.e., $g^{\mu\nu}{\cal T}_{\mu\nu}=0=u^\mu {\cal T}_{\mu\nu}$ and ${\cal T}_{\mu\nu}={\cal T}_{\nu\mu}$.

\subsection{Hydrodynamic frame ambiguity}
One notable property of nonequilibrium fluids is the ambiguity in the choice of frame, which stems from the absence of a first-principles microscopic definition of the fluid four-velocity~$u^\mu$, temperature~$T$, and chemical potential~$\mu$ out of equilibrium. In this sense, these quantities are viewed as auxiliary variables that parametrize the energy-momentum tensor and baryon current. Fixing the frame corresponds to choosing a particular definition of $(u^\mu, T, \mu)$ among infinitely many possible choices. 

The frame ambiguity can be understood through perturbative field redefinitions. Given a choice of $(u^\mu,T,\mu)$, one can redefine $u^\mu \to \tilde{u}^\mu=u^\mu +\tilde{\delta} u^\mu$, $T\to \tilde{T}=T+\tilde{\delta} T$, $\mu\to \tilde{\mu}=\mu+\tilde{\delta}\mu$, where $\tilde{\delta} u^\mu, \tilde{\delta} T, \tilde{\delta} \mu$ are of the same order as the out-of-equilibrium correction. Accordingly, $\varepsilon,p,n$ shift at the same order as the out-of-equilibrium corrections: $\varepsilon\to \tilde{\varepsilon}= \varepsilon+(\partial \varepsilon/\partial T)_{\mu} \tilde{\delta}T +(\partial \varepsilon/\partial \mu)_{T} \tilde{\delta}\mu$, $p \to \tilde{p}= p+(\partial p/\partial T)_{\mu} \tilde{\delta}T  +(\partial p/\partial \mu)_{T} \tilde{\delta}\mu$, and $n \to \tilde{n}= n+(\partial n/\partial T)_{\mu} \tilde{\delta}T  +(\partial n/\partial \mu)_{T} \tilde{\delta}\mu$. Under these redefinitions, the full energy-momentum tensor and baryon current remain invariant, while the scalar components of the out-of-equilibrium correction shift as
\begin{align}
    {\cal E}\to \tilde{\cal E}=&{\cal E} - \left[\left(\frac{\partial \varepsilon}{\partial T}\right)_{\mu} \tilde{\delta}T +\left(\frac{\partial \varepsilon}{\partial \mu}\right)_{T} \tilde{\delta}\mu \right],\\
    {\cal P} \to \tilde{\cal P}=& {\cal P}- \left[\left(\frac{\partial p}{\partial T}\right)_{\mu} \tilde{\delta}T +\left(\frac{\partial p}{\partial \mu}\right)_{T} \tilde{\delta}\mu \right] ,\\
    {\cal N}\to \tilde{\cal N}= & {\cal N}- \left[\left(\frac{\partial n}{\partial T}\right)_{\mu} \tilde{\delta}T +\left(\frac{\partial n}{\partial \mu}\right)_{T} \tilde{\delta}\mu \right],
\end{align}
and the vector components shift as
\begin{align}
   {\cal Q}_\mu\to \tilde{\cal Q}_\mu =& {\cal Q}_\mu-\left(\varepsilon+p\right) \tilde{\delta} u_\mu,\label{eq:Qmushift} \\
   {\cal J}_\mu \to \tilde{{\cal J}}_\mu =& {\cal J}_\mu - n \tilde{\delta} u_\mu, \label{eq:Jmushift} 
\end{align}
to linear order. In particular, Eqs.~\eqref{eq:Qmushift} and~\eqref{eq:Jmushift} show that one can choose a hydrodynamic frame, by setting $\tilde{\delta} u^\mu$ appropriately, in which either the heat flux or the diffusion current vanishes. 

Our framework exploits the frame ambiguity to simplify the resulting perturbation equations by absorbing the additional terms induced by the heat-flux sector into redefined perturbation variables. It is worth emphasizing that this procedure does not amount to changing the frame. Rather, for a fixed frame, we merely rewrite the perturbative system so that the heat-flux terms contribute implicitly, rather than appearing explicitly in the perturbation equations.

\section{Formulation: Even-parity sector}\label{Section:EvenParity}

\subsection{Perturbed configurations}

\subsubsection{Nonequilibrium amplitudes}
Here, we introduce fundamental variables associated with $\mathscr{S}_{\mu\nu}$ and $\mathscr{J}^\mu$, and collectively call them {\it nonequilibrium amplitudes}. To this end, we perform a harmonic decomposition for metric and fluid perturbations. In the Regge-Wheeler gauge~\cite{PhysRev.108.1063}, the metric perturbation in the Fourier domain reads
\begin{align}
    \delta g_{\mu \nu} dx^\mu dx^\nu=& -r^\ell\big[H^{\ell m} e^\nu dt^2 -2i \omega r H_1^{\ell m} dt dr \\
    &+e^\lambda H_2^{\ell m} dr^2+r^2 K^{\ell m} \Omega_{AB} dx^A dx^B\big] e^{-i \omega t} Y_{\ell m}, \nonumber
\end{align}
where $H^{\ell m},H_1^{\ell m},H_2^{\ell m},K^{\ell m}$ are functions of $r$; $Y_{\ell m}=Y_{\ell m}(\theta,\varphi)$ is a scalar spherical harmonic for a set of given integers~$(\ell,m)$, satisfying the eigenvalue equation~$[\Omega^{AB}D_A D_B+\ell(\ell+1)]Y_{\ell m} =0$, where $D_A$ is the covariant derivative associated with the unit two-sphere. Here, the normalization of $Y_{\ell m}$ is arbitrary. Note that we adopt the minus-sign convention,~$e^{-i \omega t}$, for the Fourier decomposition, and hence our definition of $\omega$ differs by an overall sign from that introduced in Refs.~\cite{Lindblom:1983ps,Detweiler:1985zz,Gualtieri:2004av}.

In the matter sector, we first decompose $\mathscr{S}_{\mu \nu}$ in terms of harmonic functions,
\begin{align}
   \mathscr{S}_{tt} =&r^{\ell-2} S_{00}^{\ell m}e^{\nu} e^{-i \omega t}Y_{\ell m},\quad
   \mathscr{S}_{tr} = r^{\ell-2} S_{01}^{\ell m} e^{-i \omega t}Y_{\ell m},\nonumber\\
   \mathscr{S}_{tA} =& r^{\ell-1}S_{0A}^{\ell m} e^{-i \omega t}E_A^{\ell m},\quad \mathscr{S}_{rr} = r^{\ell-2}S_0^{\ell m}e^\lambda e^{-i \omega t}Y_{\ell m},\nonumber\\
     \mathscr{S}_{rA} =& r^{\ell-1} S_1^{\ell m} e^{-i \omega t}E_A^{\ell m},\label{eq:EvenamplitudesOfS}\\
     \mathscr{S}_{AB} =&r^\ell e^{-i \omega t} \left(S_Z^{\ell m}Z_{AB}^{\ell m}+  S_\Omega^{\ell m}\Omega_{AB} Y_{\ell m}\right),\nonumber
\end{align}
where $S_{00}^{\ell m},S_{01}^{\ell m},S_{0A}^{\ell m},S_0^{\ell m},S_1^{\ell m},S_Z^{\ell m},S_\Omega^{\ell m}$ are functions of $r$. The even-parity vector and tensor harmonics~$E_A^{\ell m}=E_A^{\ell m}(\theta,\varphi)$ and $Z_{AB}^{\ell m}=Z_{A B}^{\ell m}(\theta,\varphi)$ are defined, respectively, by $E_A^{\ell m }=D_A Y^{\ell m}$ and $Z_{AB}^{\ell m }= [D_A D_B+(1/2)\ell(\ell+1)\Omega_{AB}]Y^{\ell m}$~\cite{Martel:2005ir}.
Likewise, we decompose $\mathscr{J}^\mu$ as
\begin{align}
\mathscr{J}^t=&-r^{\ell-2}J_0^{\ell m} e^{-i\omega t}Y_{\ell m},\quad \mathscr{J}^r=r^{\ell-2}J_1^{\ell m} e^{-i\omega t}Y_{\ell m},\nonumber\\
\mathscr{J}_A=&r^{\ell-1} J_A^{\ell m} e^{-i\omega t}E_A^{\ell m},\label{eq:EvenamplitudesOfJ}
\end{align}
where $J_0^{\ell m},J_1^{\ell m},J_A^{\ell m}$ are functions of $r$.

We call the set of the radial functions~$(S_{00}^{\ell m},S_{01}^{\ell m},S_{0A}^{\ell m},S_0^{\ell m},S_1^{\ell m},S_Z^{\ell m},S_\Omega^{\ell m},J_0^{\ell m},J_1^{\ell m},J_A^{\ell m})$ {\it nonequilibrium amplitudes}. Information of $\mathscr{S}_{\mu\nu}$ and $\mathscr{J}^\mu$ is encoded in the nonequilibrium amplitudes through
\begin{align}
    {\cal E}=&r^{\ell-2}S_{00}^{\ell m} e^{-i\omega t}Y_{\ell m},\\
    {\cal P}=&\frac{1}{3}r^{\ell-2}\left(S_0^{\ell m}+2S_\Omega^{\ell m}\right)e^{-i\omega t}Y_{\ell m},\\
    {\cal Q}_\mu=&-e^{-\nu/2} \left(0,r^{\ell-2}S_{01}^{\ell m}Y_{\ell m},r^{\ell-1}S_{0A}^{\ell m}E_A^{\ell m}\right)e^{-i \omega t},\\
    {\cal T}_{\mu\nu}=&
-\frac{1}{3}r^\ell\bigl(S_0^{\ell m}-S_\Omega^{\ell m}\bigr) e^{-i\omega t}Y_{\ell m}\nonumber\\
&\times 
\begin{pmatrix}
0 & 0 & 0 & 0 \\
0 & -2 r^{-2} e^\lambda & 0 & 0 \\
0 & 0 & \Omega_{\theta\theta} & 0 \\
0 & 0 & 0 & \Omega_{\varphi\varphi}
\end{pmatrix}
\nonumber\\
&+ r^{\ell-1} S_1^{\ell m} e^{-i \omega t}
\begin{pmatrix}
0 & 0 & 0 & 0 \\
0 & 0 & E_\theta^{\ell m} & E_\varphi^{\ell m} \\
0 & E_\theta^{\ell m} & 0 & 0 \\
0 &  E_\varphi^{\ell m}  & 0 & 0
\end{pmatrix}
\\
&+ r^\ell S_Z^{\ell m} e^{-i \omega t}
\begin{pmatrix}
0 & 0 & 0 & 0 \\
0 & 0 & 0 & 0 \\
0 & 0 & Z_{\theta\theta}^{\ell m} & Z_{\theta\varphi}^{\ell m}  \\
0 & 0 & Z_{\theta\varphi}^{\ell m}  &Z_{\varphi\varphi}^{\ell m} 
\end{pmatrix},\nonumber
\end{align}
and
\begin{align}
    {\cal N}=&-e^{\nu/2}r^{\ell-2} J_0^{\ell m} e^{-i\omega t}Y_{\ell m},\\
    {\cal J}_\mu=&\left(0,e^\lambda r^{\ell-2}J_1^{\ell m} Y_{\ell m},r^{\ell-1} J_A^{\ell m}E_A^{\ell m}\right) e^{-i\omega t}.
\end{align}

\subsubsection{Fluid perturbations}
For the fluid perturbations, we consider the displacement of $\varepsilon$, $p$, and $u^\mu$: $\varepsilon\to \varepsilon+\delta \varepsilon, p\to p+\delta p, u^\mu \to u^\mu +\delta u^\mu$. We parameterize the Lagrangian displacement of these fluid variables by two radial functions of $W^{\ell m}=W^{\ell m}(r)$ and $V^{\ell m}=V^{\ell m}(r)$ as
\begin{align}
     \xi^r=r^{\ell-1} e^{-\lambda/2} W^{\ell m} e^{-i \omega t} Y_{\ell m},\quad \xi_A=-r^\ell V^{\ell m} e^{-i \omega t} E_A^{\ell m}.
\end{align}
 We express the Eulerian perturbation of the four-velocity, $\delta u^\mu$, in terms of $W^{\ell m}$ and $V^{\ell m}$ in the following manner. First, for the Lagrangian variation~$\Delta$, using the normalization condition $(g_{\mu\nu}+\Delta g_{\mu\nu})(u^\mu+\Delta u^\mu) (u^\nu+\Delta u^\nu)=-1$ to linear order and noting $\Delta u^\mu \propto u^\mu$ linearly, we obtain 
\begin{align}
    \Delta u^\mu =\frac{1}{2} u^\mu u^\alpha u^\beta \Delta g_{\alpha \beta}.\label{eq:Deltaumu}
\end{align}
Next, using the relation,~$\Delta=\delta+{\cal L}_\xi$, where ${\cal L}_\xi$ is the Lie derivative along $\xi^\mu$, the components of $\delta u^\mu$ read
\begin{align}
    \delta u^t=&-\frac{1}{2} e^{-\nu/2} r^\ell H^{\ell m} e^{-i \omega t} Y_{\ell m},\quad  \delta u^i= - i \omega e^{-\nu/2}\xi^i.\label{eq:perturbedumu}
\end{align}
Henceforth, we omit the superscript~$\ell m$ of the radial functions.

The perturbed baryon number conservation~\eqref{eq:baryonnumberconservation} allows us to express the variation of the baryon number density in terms of the perturbation variables. Using Eqs.~\eqref{eq:Deltaumu} and~\eqref{eq:baryonnumberconservation} at linear order in perturbation leads to the fractional change of the baryon number density, ${\Delta n}/{n}=-({1}/{2})\Delta^{\mu\nu}\Delta g_{\mu \nu}+e^{\nu/2}/(i \omega n)\nabla_\mu \mathscr{J}^\mu$,
which reads
\begin{align}
       \frac{\Delta n}{n}=&r^\ell e^{-i \omega t} Y_{\ell m}\left[ \frac{H_2}{2}+K -\frac{\ell\left(\ell+1\right)}{r^2}V\right.\nonumber\\
    &\left.-e^{-\lambda/2}\frac{(\ell+1)W+ rW'}{r^2}+\mathscr{C}_J\right],\label{eq:Deltann}
\end{align}
where $\mathscr{C}_J$ is a current-induced contribution, defined by
\begin{align}
    \mathscr{C}_J:= &\frac{e^{\nu/2}}{nr^2}\left[J_0+\frac{1}{i\omega r}\left\{\left(\ell+4\pi \left(\varepsilon+p\right) r^2 e^\lambda\right)J_1\right.\right.\nonumber\\
    &\left.\left.+r J_1'-\ell\left(\ell+1\right)J_A\right\}\right].\label{eq:CJ}
\end{align}
It follows that $\mathscr{J}^\mu$ contributes to variations in the baryon number density.

Variations of the thermodynamic quantities are expressed in terms of the perturbation variables as follows. First, let us consider the projection of the perturbed conservation law, i.e., the second equation in Eq.~\eqref{eq:fieldeqs}, onto the direction along the fluid four-velocity, together with Eq.~\eqref{eq:baryonnumberconservation}, up to linear order, yielding
\begin{align}
    \Delta \varepsilon=\left(\varepsilon+p\right)\frac{\Delta n}{n}-\frac{e^{\nu/2}}{i \omega}\left(u^\mu \nabla_\nu \mathscr{S}^\nu_{~\mu}+\frac{\varepsilon+p}{n}\nabla_\mu \mathscr{J^\mu}\right).\label{eq:Deltaepsilon}
\end{align}
This relation can be understood in terms of thermodynamics as follows. In general, the first law of thermodynamics reads
\begin{align}
\Delta\varepsilon=\left(\varepsilon+p\right)\frac{\Delta n}{n}+n T \Delta s+n \mu_i\Delta Y^i,\label{eq:firstlaw}
\end{align}
where $T$ and $s$ denote the temperature and the entropy per baryon, respectively. The variables~$Y^i$ denote the composition fractions chosen as a set of independent thermodynamic variables, and $\mu^i$ correspond to the quantities conjugate to $Y^i$. In general, $\mu^i$ may take the form of a combination of the chemical potentials of individual particle species.\footnote{A useful example is provided in Appendix~B of Ref.~\cite{Gualtieri:2004av}. In this example, first the lepton fraction~$Y_L$ is chosen as an independent variable. Initially there are the fractions to the baryon number for electrons~$Y_e$, protons~$Y_p$, neutrons~$Y_n$, neutrinos~$Y_\nu$, as well as the corresponding chemical potentials~$\mu_e$, $\mu_p$, $\mu_n$, $\mu_\nu$. The combination of these variables is then rewritten by using the baryon-fraction constraint~$Y_p+Y_n=1$ and the electron-capture reaction under chemical equilibrium, which gives $dY_e=dY_p$ and $\mu_e+\mu_p=\mu_n+\mu_\nu$. These relations yield $\mu_e dY_e+\mu_p dY_p+\mu_n dY_n+\mu_\nu dY_\nu=\mu_\nu dY_L$, where $dY_L = dY_e+dY_\nu$. Thus, $\mu_\nu$ can be viewed as being conjugate to $Y_L$ in this example.} Comparison of Eq.~\eqref{eq:firstlaw} with Eq.~\eqref{eq:Deltaepsilon} reveals that the out-of-equilibrium contributions in Eq.~\eqref{eq:Deltaepsilon}, i.e., the flux~$\mathscr{S}_t^{~\mu}$ and the current~$\mathscr{J}^\mu$, induce the combination of entropy and composition perturbations:
\begin{align}
    n T \Delta s+n\mu_i\Delta Y^i=&-\frac{e^{\nu/2}}{i \omega}\left(u^\mu \nabla_\nu \mathscr{S}^\nu_{~\mu}+\frac{\varepsilon+p}{n}\nabla_\mu \mathscr{J^\mu}\right),\nonumber\\
   =&-r^{\ell-2} e^{-i \omega t}Y_{\ell m}\left[\mathscr{C}_S+r^2\left(\varepsilon+p\right)\mathscr{C}_J\right],\label{eq:NonAdiabaticCondition}
\end{align}
where we have introduced a flux-induced contribution,
\begin{align}
    \mathscr{C}_S:=&S_{00}+\frac{1}{i \omega r}\left[\left\{1+\left(\ell-1\right)e^{-\lambda} -4\pi r^2 \left(\varepsilon-p\right) \right\}S_{01}\right.\nonumber\\
    &\left.+e^{-\lambda} r S_{01}'-\ell\left(\ell+1\right)S_{0A}\right].\label{eq:CS}
\end{align}
It is worth noting that the first law of thermodynamics provides no information on how the out-of-equilibrium effects are distributed between entropy and chemical-composition variations. Additional closure relations, such as evolution equations for entropy and/or matter species, are required to determine $\Delta s$ and $\Delta Y^i$ separately, under the constraint imposed by Eq.~\eqref{eq:NonAdiabaticCondition}.\footnote{See also Refs.~\cite{Camelio:2022ljs,Camelio:2022fds} for the formulation and numerical comparison, respectively, of a reacting multicomponent fluid and two effective bulk-stress models.} These relations are distinct from the constitutive relations specifying $\mathscr{S}_{\mu\nu}$ and $\mathscr{J}^\mu$.

The pressure perturbation is connected to the other thermodynamic quantities through the equation of state~$\varepsilon=\varepsilon(p,s,Y^i)$. To this end, we parameterize the Lagrangian perturbation of pressure by
\begin{align}
   \Delta p=c_s^2 \left(\Delta \varepsilon-n T\alpha_1 \Delta s-n \mu_i \alpha_{2i}\Delta Y^i\right),\label{eq:thermodynamicalrelation}
\end{align}
where $c_s^2:=(\partial p/\partial \varepsilon)_{s,Y^i}$ is the adiabatic speed of sound squared. The coefficients~$\alpha_1:=(nT)^{-1}(\partial \varepsilon/\partial s)_{p,Y^i}$ and $\alpha_{2i}:=(n \mu^i)^{-1}(\partial \varepsilon/\partial Y^i)_{p,s}$ are determined by the equation of state, whereas $\Delta s$ and $\Delta Y^i$ must be specified separately through evolution equations or additional assumptions, as noted above. Given these relations, one can obtain the Eulerian variation in the energy density and pressure via
\begin{align}
    \delta \varepsilon=& \Delta\varepsilon -\xi^r \varepsilon',\quad \delta p= \Delta p-\xi^r p'.
\end{align}

\subsection{Variable redefinition}
The frame ambiguity in non-equilibrium fluids allows us to simplify the resulting perturbation equations by introducing appropriate variables. Indeed, as will be seen below, the flux components, i.e., ~$S_{01}$ and $S_{0A}$, do not appear in the perturbation equations explicitly but are absorbed into redefined perturbation variables and contribute implicitly. Therefore, the resulting system reduces to a system close to the Lindblom-Detweiler equations with a modest number of additional terms.

Redefinition of variables is performed as follows.
We first redefine the functions, $W$ and $V$, by
\begin{align}
    \hat{W}=&W+\frac{e^{-\lambda/2}}{i\omega r \left(\varepsilon+p\right)}S_{01},\quad
\hat{V}=V-\frac{1}{i \omega r \left(\varepsilon+p\right)}S_{0A}.\label{eq:hatWhatV}
\end{align}
This operation is schematically equivalent to a hydrodynamic frame transformation~$\delta u^\mu\to \delta u^\mu+{\cal Q}^\mu/(\varepsilon+p)$ through Eq.~\eqref{eq:perturbedumu}. Moreover, we introduce auxiliary radial functions~$X(r)$, $E(r)$, $\Sigma(r)$, and ${\cal Y}^i(r)$:
\begin{align}
    X Y_{\ell m}e^{-i \omega t}=&-e^{\nu/2}r^{-\ell} \Delta p,\label{eq:X}\\
     E Y_{\ell m}e^{-i \omega t}=&-e^{\nu/2}r^{-\ell} \Delta \varepsilon,\label{eq:E}\\
      \Sigma Y_{\ell m}e^{-i \omega t}=&-e^{\nu/2}r^{-\ell} \Delta s,\label{eq:Sigma}\\
       {\cal Y}^i Y_{\ell m}e^{-i \omega t}=&-e^{\nu/2}r^{-\ell}\Delta Y^i.\label{eq:Y}
\end{align}
Then, we redefine these as 
\begin{align}
    \hat{X}  =&X-\frac{e^{-\lambda+\nu/2}p'}{i\omega r^2 \left(\varepsilon+p\right)}S_{01},\label{eq:hatX}\\
     \hat{E} =&E-\frac{e^{-\lambda+\nu/2}\varepsilon'}{i\omega r^2 \left(\varepsilon+p\right)}S_{01},\label{eq:hatE}\\
      \hat{\Sigma} =&\Sigma -\frac{e^{-\lambda+\nu/2}s'}{i\omega r^2 \left(\varepsilon+p\right)}S_{01},\label{eq:hatSigma}\\
       \hat{\cal Y}^i =&{\cal Y}^i-\frac{e^{-\lambda+\nu/2}Y^{i'}}{i\omega r^2 \left(\varepsilon+p\right)}S_{01}.\label{eq:hatY}
\end{align}
In terms of these redefined variables, the thermodynamic relation equivalent to Eq.~\eqref{eq:thermodynamicalrelation} holds
\begin{align}
\hat{E}=c_s^{-2} \hat{X}+ n T \alpha_1 \hat{\Sigma}+n \mu_i \alpha_{2i}\hat{\cal Y}^i.\label{eq:thermodynamicalEXSigma}
\end{align}

\subsection{Perturbation equations}\label{Section:PerturbationEqs}
Now, we derive perturbation equations by combining the components of Eq.~\eqref{eq:fieldeqs} following the manner outlined in Appendix~\ref{Appendix:Derivation}. First of all,  $\delta G_{\theta\varphi}=8\pi (\delta T_{\theta\varphi}+\mathscr{S}_{\theta\varphi})$ yields
\begin{align}
    H_2=H+16 \pi S_Z,\label{eq:eqforH2}
\end{align}
which determines $H_2$ by $H$ and $S_Z$. We then find the following four differential equations,
\begin{widetext}
\begin{align}
H_1'=&\frac{e^\lambda}{r} H+\left[4\pi r \left(\varepsilon-p\right)e^\lambda -\frac{2m e^\lambda}{r^2}-\frac{\ell+1}{r}\right]H_1+\frac{e^\lambda}{r} K-16 \pi\frac{\varepsilon+p}{r}e^\lambda \hat{V}+16\pi\frac{e^\lambda}{r}S_Z,\label{eq:eqforH1}\\
K'=&\frac{H}{r}+\frac{\ell\left(\ell+1\right)}{2r}H_1+\left(\frac{\nu'}{2}-\frac{\ell+1}{r}\right)K-8\pi\frac{\varepsilon+p}{r}e^{\lambda/2}\hat{W}+\frac{16\pi}{r}S_Z,\label{eq:eqforK}\\
\hat{W}'=&\frac{r}{2}e^{\lambda/2}H+r e^{\lambda/2}K+\frac{r e^{(\lambda-\nu)/2}}{\varepsilon+p}\hat{E}-\frac{\ell+1}{r}\hat{W}-\frac{\ell\left(\ell+1\right)}{r}e^{\lambda/2}\hat{V}-\frac{e^{\lambda/2}}{r\left(\varepsilon+p\right)}S_{00}
+8\pi r e^{\lambda/2} S_Z,\label{eq:eqforW}\\
\hat{X}'=&\frac{\varepsilon+p}{2}e^{\nu/2}\left(\frac{1}{r}-\frac{\nu'}{2}\right)H+\frac{\varepsilon+p}{2}e^{\nu/2}\left[\omega^2 r e^{-\nu}+\frac{\ell\left(\ell+1\right)}{2r}\right]H_1+\frac{\varepsilon+p}{2}e^{\nu/2}\left(\frac{3\nu'}{2}-\frac{1}{r}\right)K\nonumber\\
&-\frac{\ell}{r}\hat{X}-\frac{\varepsilon+p}{r}e^{(\lambda+\nu)/2}\left[\omega^2 e^{-\nu}+\frac{e^{-\lambda}}{4r^2}\left\{e^{2\lambda}\left(1+8\pi r^2 p \right)^2+2e^\lambda\left(3+8\pi r^2 p \right)-7\right\}\right]\hat{W}+\frac{\ell\left(\ell+1\right)}{r^2}e^{\nu/2}p' \hat{V}\label{eq:eqforpX}\\
&+\frac{e^{\nu/2}}{2r^3}\left[2\ell-1+e^\lambda \left(1+8\pi r^2 p\right)\right]S_0+\frac{e^{\nu/2}}{r^2}S_0'-\frac{e^{\nu/2}}{r^3}\left[\ell(\ell+1)+8\pi r^2(\varepsilon+p)\right]S_1-\frac{2 e^{\nu/2}}{r^3}S_\Omega,\nonumber
\end{align}
and the following two relations among the perturbation variables, 
\begin{align}
    &\left[3m+\frac{\left(\ell+2\right)\left(\ell-1\right)}{2}r+4\pi r^3 p \right]H=-\left[\frac{\ell\left(\ell+1\right)}{2}\left(m+4\pi r^3 p\right) -\omega^2 r^3 e^{-\lambda-\nu}\right]H_1\nonumber\\
    &+\left[\frac{\left(\ell+2\right)\left(\ell-1\right)}{2}r-\omega^2 r^3 e^{-\nu}-\frac{e^\lambda}{r}\left(m+4\pi r^3 p \right)\left(3m-r+4\pi r^3 p \right)\right]K+8 \pi r^3e^{-\nu/2}\hat{X}-8\pi r S_0-16 \pi r e^{-\lambda}S_1,\label{eq:eqforH}\\
    &\hat{X}=\frac{\varepsilon+p}{2}e^{\nu/2}H-\frac{p'}{r}e^{(\nu-\lambda)/2}\hat{W}+\omega^2 \left(\varepsilon+p\right)e^{-\nu/2}\hat{V}\nonumber\\
    &+\frac{e^{-\lambda+\nu/2}}{r^2}\left[\ell+e^\lambda \left\{1-4\pi r^2 \left(\varepsilon-p\right)\right\}\right]S_1+\frac{e^{-\lambda+\nu/2}}{r}S_1'-\frac{\ell^2+\ell-2}{2r^2}e^{\nu/2}S_Z+\frac{e^{\nu/2}}{r^2}S_\Omega.\label{eq:eqforX}
\end{align}
Moreover, the out-of-equilibrium corrections~$\mathscr{S}_{\mu\nu}$ satisfy
\begin{align}
n e^{-\nu/2}r^2 \left[T\left(\hat{\Sigma}+\frac{e^{-\lambda+\nu/2}s'}{i\omega r^2\left(\varepsilon+p\right)}S_{01}\right)+\mu_i\left(\hat{\cal Y}^i+\frac{e^{-\lambda+\nu/2}{Y^i}'}{i\omega r^2\left(\varepsilon+p\right)}S_{01}\right)\right]= \mathscr{C}_S+r^2\left(\varepsilon+p\right)\mathscr{C}_J, \label{eq:eqforSigmaYi}
\end{align}
\end{widetext}
where $\mathscr{C}_S$ and $\mathscr{C}_J$ are defined in Eqs.~\eqref{eq:CS} and~\eqref{eq:CJ}, respectively. We provide the expressions for these equations online~\cite{OnlineLink}.

\subsection{Structure}
Let us examine the structure of the perturbation equations presented above. There are nine equations, Eqs.~\eqref{eq:thermodynamicalEXSigma}--\eqref{eq:eqforSigmaYi}, for the $(9+i)$ variables~$(H,H_1,H_2,K,\hat{W},\hat{V},\hat{X},\hat{E},\hat{\Sigma},\hat{\cal Y}^i)$ with given corrections~$\mathscr{S}_{\mu\nu}$ and~$\mathscr{J}^\mu$ encapsulated in the set of the nonequilibrium amplitudes through Eqs.~\eqref{eq:EvenamplitudesOfS} and~\eqref{eq:EvenamplitudesOfJ}. In the general case, these equations do not fix the decomposition of the non-adiabatic perturbations into $\hat{\Sigma}$ and $\hat{\cal Y}^i$. Additional $i$ closure relations are required to close the system. It is worth emphasizing that our formalism provides perturbation equations for the generic out-of-equilibrium corrections given in Eqs.~\eqref{eq:fieldeqs} and~\eqref{eq:baryonnumberconservation} but does not itself provide additional evolution or constraint equations for independent dynamical variables that may arise in particular fluid models, such as M{\"u}ller-Israel-Stewart fluids~\cite{1967ZPhy..198..329M,Israel:1976tn,Israel:1976efz,1977RSPSA.357...59S,1979RSPSA.365...43I,Israel:1979wp}. The system presented above can be directly applied without introducing further independent variables to fluid models in which the out-of-equilibrium corrections are specified by constitutive relations expressed in the ordinary hydrodynamic variables and their derivatives, as in the BDNK fluid~\cite{Bemfica:2017wps,Bemfica:2019knx,Kovtun:2019hdm,Hoult:2020eho,Bemfica:2020zjp}. It is worth noting that, depending on the structure of the nonequilibrium amplitudes for a given fluid model, the unhatted variables, such as $W$, $V$, $\Sigma$, and ${\cal Y}^i$, may be more convenient for practical implementations. Indeed, as discussed in Appendix~\ref{Appendix:BDNK}, Eq.~\eqref{eq:eqforSigmaYi} is easier to solve in the BDNK system when expressed in terms of $W$, $V$, $\Sigma$, and ${\cal Y}^i$.

Let us discuss below a few restricted cases and consistency with literature:
\begin{itemize}
    \item 
In the absence of out-of-equilibrium corrections, i.e.,~$\mathscr{S}_{\mu\nu}=0=\mathscr{J}^\mu$, the above system reduces to the Lindblom-Detweiler formalism~\cite{Lindblom:1983ps,Detweiler:1985zz}.\footnote{\label{footnote:OmegaConvention}Note the conventional difference in the sign of $\omega$; the same convention is adopted in Ref.~\cite{Gualtieri:2004av}.} 

\item
In the absence of the flux components~$\mathscr{S}_{t\mu}$ and $\mathscr{J}^\mu$, i.e., $S_{00}=S_{01}=S_{0A}=J_0=J_1=J_A=0$, the perturbation remains adiabatic, as noted in Eq.~\eqref{eq:NonAdiabaticCondition}. In such a case, there are eight perturbation equations, Eqs.~\eqref{eq:thermodynamicalEXSigma}--\eqref{eq:eqforX}, for eight variables, due to $\hat{\Sigma}=\hat{\cal Y}^i=0$. This class includes the Landau-Lifshitz frame fluid with only shear and bulk viscosity adopted in Refs.~\cite{HegadeKR:2024agt,Saketh:2024juq}. 

\item In the presence of heat flux ($S_{01}$ and $S_{0A}$), our formalism recovers Eqs.~(5.2)--(5.4) and Eqs.~(5.6)--(5.9), as well as Eq.~(5.11) in Ref.~\cite{Gualtieri:2004av} with the choice of $(S_{01},S_{0A})=- \kappa r (Q_1,Q_2)$, following the notation therein~(see also footnote~\ref{footnote:OmegaConvention}).\footnote{Equations~(5.5) and~(5.10) therein arise from the (truncated) M{\"u}ller-Israel-Stewart equations.} When lepton-fraction perturbations are included, Eqs.~\eqref{eq:thermodynamicalEXSigma} and~\eqref{eq:eqforSigmaYi} reproduce Eqs.~(B30) and~(B33) in the reference, whereas the remaining equations therein require additional transport closure relations. 
\end{itemize}

An intriguing feature of the system is that the contribution of the flux components of $\mathscr{S}^{\mu\nu}$, encoded in $S_{01}$ and $S_{0A}$, can be absorbed into the fluid variables, i.e., $\hat{X}, \hat{E}, \hat{\Sigma}, \hat{\cal Y}^i, \hat{W}, \hat{V}$, through an appropriate redefinition. As a result, the flux contributions do not appear in Eqs.~\eqref{eq:thermodynamicalEXSigma}--\eqref{eq:eqforX} explicitly, and hence, the system retains a form close to the Lindblom-Detweiler formalism with a modest number of additional terms induced by the out-of-equilibrium energy density, pressure, and shear stress. This is a clear manifestation of the frame ambiguity: by redefining the fluid variables, one can absorb the heat-flux sector, so that its effect is encoded in the perturbation equations implicitly, rather than appearing explicitly. It is worth emphasizing that our procedure does not amount to changing of hydrodynamic frame; rather, for a fixed frame, it merely rewrites the perturbation equations in a form in which the heat-flux sector does not appear explicitly.

It is worth noting that, once the nonequilibrium amplitudes are specified through constitutive relations, there are two distinct ways of treating the resulting system, depending on whether or not the perturbation variables are expanded in transport coefficients. On the one hand, with such an expansion, the problem can be equivalently viewed as an inhomogeneous problem in which the nonequilibrium amplitudes serve as source terms driving corrections to the perfect-fluid quasinormal modes, as discussed in Section~\ref{Section:Application}. On the other hand, without such an expansion, the system may have a higher differential order than in the perfect-fluid case, if the nonequilibrium amplitudes depend on derivatives of the perturbation variables. For instance, Eq.~\eqref{eq:eqforpX} contains $S_0'$ that can lead to second-order derivatives of the variables. In addition, Eqs.~\eqref{eq:eqforH2},~\eqref{eq:eqforH} and~\eqref{eq:eqforX} can no longer be algebraic relations. These observations indicate that the solution space of perturbations for non-equilibrium fluids is effectively enlarged, thereby potentially leading to branches that have no regular perfect-fluid counterpart in the limit of vanishing nonequilibrium contributions. Physically, such branches would correspond to new families of quasi-normal modes in characteristic oscillations. In Section~\ref{Section:Application}, we discuss the possible feature leading to new mode families for the BDNK fluid.

In Appendix~\ref{Appendix:BDNK}, we provide the nonequilibrium amplitudes for BDNK fluids. It is useful to clarify the relation between our framework specialized to the BDNK fluid and the framework constructed in Ref.~\cite{Redondo-Yuste:2024vdb}. In Ref.~\cite{Redondo-Yuste:2024vdb}, thermal conductivity is set to zero and the matter sector is restricted to be barotropic up to the level of perturbations. Thus, non-adiabatic thermodynamic degrees of freedom, such as entropy or composition perturbations, are effectively frozen. Our framework, by contrast, keeps the non-barotropic sector explicit, allowing entropy and composition perturbations, as well as thermal conductivity to enter perturbations. Moreover, since Ref.~\cite{Redondo-Yuste:2024vdb} does not introduce a conserved baryon current, while our formulation does, the two matter sectors are not identical, and hence a direct comparison is non-trivial. Specifically, Ref.~\cite{Redondo-Yuste:2024vdb} retains the energy-density perturbation as an independent dynamical variable without using the baryon number conservation. In our formulation, by contrast, the conservation laws of energy and momentum of the fluid and the baryon current are instead combined to eliminate the energy-density perturbation~(see Eq.~\eqref{eq:Deltaepsilon}).

\section{Formulation: Odd-parity sector}\label{Section:OddParity}

\subsection{Perturbed configuration}
In the Regge-Wheeler gauge~\cite{PhysRev.108.1063}, the metric fluctuation reads
\begin{align}
    \delta g_{\mu \nu} dx^\mu dx^\nu=\left(2 h^{\ell m} B_A^{\ell m} dt dx^A+2 h_1^{\ell m} B_A^{\ell m}  dr dx^A\right) e^{-i \omega t},
\end{align}
where $(h^{\ell m}, h_1^{\ell m})=(h^{\ell m}(r), h_1^{\ell m}(r))$; $B_A^{\ell m}=B_A^{\ell m}(\theta,\varphi)$ is the odd-parity vector spherical harmonic, defined by $B_A^{\ell m}=-\varepsilon_A^{~B}D_B Y^{\ell m}$ with $\varepsilon_{AB}$ being the Levi-Civita tensor on the unit two-sphere, whose nonvanishing component is $\varepsilon_{\theta \varphi}=\sin\theta$~\cite{Martel:2005ir}. In the odd-parity sector, the thermodynamic scalar quantities are not perturbed. The Eulerian displacement of the fluid four-velocity~$\delta u^\mu$ is parametrized by a radial function~$U_{\ell m}=U_{\ell m}(r)$ as 
\begin{align}
    \delta u^\mu=\frac{U_{\ell m} e^{\nu/2}}{4\pi r^2 \left(\varepsilon+p\right)} e^{-i \omega t}\left(0,0,B_\theta^{\ell m},\frac{B_\varphi^{\ell m}}{\sin^2\theta}\right).\label{eq:delau}
\end{align}

As in the even-parity sector, we introduce nonequilibrium amplitudes by decomposing $\mathscr{S}_{\mu \nu}$ in terms of harmonic functions,
\begin{align}
    \mathscr{S}_{tA}=& S_{0A}^{-,\ell m}e^{-i \omega t} B_A^{\ell m},\quad \mathscr{S}_{rA}= S_{1A}^{-,\ell m}e^{-i \omega t} B_A^{\ell m},\label{eq:OddamplitudesOfS}\\
    \mathscr{S}_{AB}=&S_Z^{-,\ell m} e^{-i \omega t}Z_{AB}^{-,\ell m},\nonumber
\end{align}
where $(S_{0A}^{-,\ell m},S_{1A}^{-,\ell m},S_{Z}^{-,\ell m})$ are functions of $r$; $Z_{AB}^{-,\ell m}$ is the odd-parity tensor spherical harmonic, defined by $Z_{AB}^{-,\ell m}=-(1/2)(\varepsilon_A^{~C} D_B+\varepsilon_B^{~C}D_A)D_C Y^{\ell m}$~\cite{Martel:2005ir}. Moreover, $\mathscr{J}^\mu$ is decomposed as
\begin{align}
    \mathscr{J}_A=&r^2J_A^{-,\ell m}e^{-i \omega t} B_A^{\ell m},\label{eq:OddamplitudesOfJ}
\end{align}
where $J_A^{-,\ell m}$ is a function of $r$. The functions $(S_{0A}^{-,\ell m},S_{1A}^{-,\ell m},S_{Z}^{-,\ell m},J_A^{-,\ell m})$ are referred to as the odd-parity nonequilibrium amplitudes.

In the odd-parity sector, scalar quantities, such as ${\cal E}$, ${\cal P}$, and ${\cal N}$ in Eqs.~\eqref{eq:Smunu} and~\eqref{eq:Jmu}, are not induced out of equilibrium. One can express 
\begin{align}
    {\cal Q}_\mu=&-e^{-\nu/2}\left(0,0,S_{0A}^{-,\ell m}B_A^{\ell m}\right)e^{-i\omega t},\\
    {\cal T}_{\mu \nu}=&
S_{1A}^{-,\ell m} e^{-i \omega t}
\begin{pmatrix}
0 & 0 & 0 & 0 \\
0 & 0 & B_\theta^{\ell m} & B_\varphi^{\ell m} \\
0 & B_\theta^{\ell m} & 0 & 0 \\
0 &  B_\varphi^{\ell m}  & 0 & 0
\end{pmatrix}
\\
&+ S_Z^{-,\ell m} e^{-i \omega t}
\begin{pmatrix}
0 & 0 & 0 & 0 \\
0 & 0 & 0 & 0 \\
0 & 0 & Z_{\theta\theta}^{-,\ell m} & Z_{\theta\varphi}^{-,\ell m}  \\
0 & 0 & Z_{\theta\varphi}^{-,\ell m}  &Z_{\varphi\varphi}^{-,\ell m} 
\end{pmatrix},\nonumber
\end{align}
and
\begin{align}
    {\cal J}_\mu=&\left(0,0,r^2 J_A^{-,\ell m}B_A^{\ell m}\right) e^{-i\omega t}.
\end{align}
Hereafter, we omit $\ell m$.

\subsection{Perturbation equations}
The out-of-equilibrium corrections~$\mathscr{S}_{\mu\nu}$ satisfy Eq.~\eqref{eq:fieldeqs}. First, $\delta G_{r\theta}=8\pi(\delta T_{r\theta}+\mathscr{S}_{r\theta})$ allows us to express $h_1$ as
\begin{align}
    h_1=-\frac{r^2}{\Lambda_\ell e^\nu-\omega^2 r^2}\left[i \omega \left(h'-\frac{2h}{r}\right)-16\pi e^\nu S_{1A}^- \right],\label{eq:Eqforh1}
\end{align}
where $\Lambda_\ell=\ell^2+\ell-2$. With this expression, $\delta G_{\theta \theta}=8\pi(\delta T_{\theta \theta}+\mathscr{S}_{\theta \theta})$ yields the expression,~$h_1'=h_1'(h',h)$. Substituting it into $\delta (\nabla_\mu T^{\mu \theta})=-\nabla_\mu \mathscr{S}^{\mu \theta}$, one finds
\begin{align}
    \hat{U}=&-4\pi e^{-\nu} \left(\varepsilon+p\right) h\nonumber\\
    &+\frac{4\pi e^{-\lambda}}{i\omega r} \left[1+e^\lambda\left\{1-4\pi r^2\left(\varepsilon-p\right)\right\}\right]S_{1A}^-\label{eq:EqforU}\\
    &+\frac{4\pi e^{-\lambda}}{i\omega}{S_{1A}^-}'- \frac{2\pi \Lambda_\ell}{i\omega r^2}  S_Z^-,\nonumber
\end{align}
where we have introduced
\begin{align}
    \hat{U}=U-4\pi e^{-\nu} S_{0A}^-.
\end{align}
Lastly, using these results, we obtain from $\delta G_{t \theta}=8\pi(\delta T_{t \theta}+\mathscr{S}_{t \theta})$ that
\begin{widetext}
\begin{align}
&\left(\Lambda_\ell e^\nu-\omega^2r^2\right)h''=\left[4\pi\Lambda_\ell  r e^{\lambda+\nu}\left(\varepsilon+p\right)-r\omega^2\left\{e^\lambda\left(4\pi r^2\left(\varepsilon-p\right)-1\right)+3\right\} \right]h'\nonumber\\
&+\frac{1}{r^2}\left[2\left(\Lambda_\ell e^\nu+2\omega^2 r^2\right)+e^{\lambda-\nu}\left\{\Lambda_\ell e^{2\nu}\left(\Lambda_\ell -8\pi r^2\left(\varepsilon+p\right)\right)-2\omega^2 r^2 e^\nu\left(\Lambda_\ell +1-4\pi r^2 \left(\varepsilon-p\right)\right) +\omega^4r^4\right\} \right] h\label{eq:Eqforh}\\
&+\frac{16\pi e^\nu}{i \omega r}\left[\Lambda_\ell e^\nu+e^\lambda\left\{\Lambda_\ell e^\nu\left(1-4\pi r^2\left(\varepsilon-p\right)\right)-2\omega^2 r^2\left(1-2\pi r^2\left(\varepsilon-3p\right)\right)\right\} +2 \omega^2 r^2\right]S_{1A}^-\nonumber\\
&+\frac{16\pi e^\nu}{i \omega}\left(\Lambda_\ell e^\nu-\omega^2 r^2\right) {S_{1A}^-}'-\frac{8\pi e^\lambda \left(\Lambda_\ell e^\nu -\omega^2r^2\right)^2}{i \omega r^2}S_Z^-.\nonumber
\end{align}
\end{widetext}

Now, we reduce the system to a more familiar and simpler form by combining Eqs.~\eqref{eq:Eqforh1} and \eqref{eq:Eqforh}. First, we introduce an auxiliary variable~$\psi=\psi(r)$ so that
\begin{align}
   h_1=&e^{-\left(\nu-\lambda\right)/2}r \psi.\label{eq:h1inpsiandchis}
\end{align}
We then find
\begin{widetext}
\begin{align}
    &e^{\left(\nu-\lambda\right)/2}\left(e^{\left(\nu-\lambda\right)/2} \psi'\right)'+\left[\omega^2-e^\nu \left(\frac{\ell\left(\ell+1\right)}{r^2}-\frac{6m}{r^3}+4\pi \left(\varepsilon-p\right)\right)\right]\psi\nonumber\\
    &=-\frac{8\pi}{r^2}e^{\left(3\nu-\lambda\right)/2}\left[2r S_{1A}^--\left\{e^\lambda \left(1+8\pi r^2p \right)-3\right\}S_Z^--r {S_Z^-}'\right],\label{eq:RWinterior}
\end{align}
and 
\begin{align}
    \hat{U}=&\frac{4\pi e^{-\left(\nu+\lambda\right)/2}}{i \omega} \left(\varepsilon+p\right) \left(r \psi\right)' 
    +\frac{4\pi e^{-\lambda}}{i\omega r} \left[1+e^\lambda\left\{1-4\pi r^2\left(\varepsilon-p\right)\right\}\right]S_{1A}^-
    +\frac{4\pi e^{-\lambda}}{i\omega}{S_{1A}^-}' -\frac{2\pi}{i\omega r^2} \left[ \Lambda_\ell+16\pi r^2 \left(\varepsilon+p\right)\right] S_Z^-,\label{eq:EqforUinpsi}
\end{align}
\end{widetext}
with
\begin{align}
    h=& -\frac{e^{\left(\nu-\lambda\right)/2}}{i \omega}\left(r \psi\right)'+\frac{8\pi}{i \omega}e^\nu S_Z^-.\label{eq:hinpsiandchis}
\end{align}
The expressions for these equations are provided online~\cite{OnlineLink}.

\subsection{Structure}
For three perturbation variables~$(h,h_1,U)$, there are three equations~\eqref{eq:Eqforh1},~\eqref{eq:EqforU}, and \eqref{eq:Eqforh}, with the nonequilibrium amplitudes. In the absence of the nonequilibrium contributions, Eqs.~\eqref{eq:Eqforh1},~\eqref{eq:EqforU}, and \eqref{eq:Eqforh} reduce to Eqs.~(B30)--(B32) in Ref.~\cite{Katagiri:2025qze}. The static limit of the perfect-fluid part recovers Eq.~(13) in Ref.~\cite{Pani:2018inf}. The resulting ``master'' equations are presented in Eqs.~\eqref{eq:RWinterior} and~\eqref{eq:EqforUinpsi}. Equation~\eqref{eq:RWinterior} reduces to the Regge-Wheeler equation in the exterior~\cite{PhysRev.108.1063}. Moreover, our framework recovers the two coupled wave equations derived in Ref.~\cite{Redondo-Yuste:2024vdb} for the BDNK fluid.

Without out-of-equilibrium contributions, Eq.~\eqref{eq:Eqforh} reduces to a differential equation for $h$, decoupled from $h_1$ or $U$. Once $h$ is solved, one can obtain $h_1$ and $U$ from Eqs.~\eqref{eq:Eqforh1} and~\eqref{eq:EqforU}. In particular, $U$ is algebraically determined in terms of $h$, and is therefore not an independent dynamical variable for stellar oscillations.

By contrast, out-of-equilibrium contributions introduce couplings among $h$, $h_1$, and $U$. In particular, since the nonequilibrium amplitudes may depend not only on $h$ and $h_1$ but also on their derivatives as well as $U'$ and $U''$, Eq.~\eqref{eq:EqforUinpsi} can be viewed as a differential equation for $U$. This indicates that $U$ can become an independent contribution to stellar oscillations, potentially giving rise to new families of quasi-normal modes in characteristic oscillations. In Section~\ref{Section:Application}, we discuss the new mode family for the BDNK fluid.

The odd-parity sector is considerably simpler than the even-parity sector, owing to the absence of variations in scalar quantities associated with fluid and thermodynamical fluctuations. In particular, only the shear~${\cal T}_{\mu\nu}$, heat-flux~${\cal Q}^\mu$, and diffusion-current~${\cal J}^\mu$ contributions can affect stellar oscillations. Moreover, Eq.~\eqref{eq:EqforU} implies that the flux part~$S_{0A}^-$ can be absorbed into a redefined $U$ through $\delta u^\mu\to \delta u^\mu+{\cal Q}^\mu/(\varepsilon+p)$. 

It is worth noting that the perturbed baryon number conservation law~\eqref{eq:baryonnumberconservation} identically holds, and hence, $\mathscr{J}^\mu$ does not give rise to an independent equation. This implies that $J_A^-$ is not constrained by the baryon-number conservation.

\section{Application: BDNK fluids}
\label{Section:Application}

We now apply the generic formalism developed in the previous sections to the BDNK fluid and discuss the structure of the perturbation equations, characteristic mode shifts, as well as the possible structural feature that might give rise to new mode families.

\subsection{Basics of BDNK fluids}

\subsubsection{Constitutive relations}
The BDNK fluid is parametrized by the following constitutive relations:~\cite{Bemfica:2017wps,Bemfica:2019knx,Kovtun:2019hdm,Hoult:2020eho,Bemfica:2020zjp}
\begin{align}
    {\cal E}=&\tau_{\cal E}\left[u^\mu \nabla_\mu \varepsilon+\left(\varepsilon+p\right)\Theta\right],\label{eq:BDNKtauE}\\
    {\cal P}=& -\zeta \Theta+\tau_{\cal P}\left[u^\mu \nabla_\mu \varepsilon+\left(\varepsilon+p\right)\Theta\right],\label{eq:BDNKtauP}\\
    {\cal Q}^\mu=& \kappa T \frac{\varepsilon+p}{n}\Delta^{\mu\sigma}\nabla_\sigma \left(\mu/T\right)\label{eq:BDNKkappa_tauQ}\\
    &+\tau_{\cal Q}\left[\left(\varepsilon+p\right)u^\nu \nabla_\nu u^\mu +\Delta^{\mu \sigma} \nabla_\sigma p\right],\nonumber\\
    {\cal T}^{\mu\nu}=&-2\eta \sigma^{\mu\nu},\\
    {\cal N}=&0,\\
    {\cal J}^\mu=&0,
\end{align}
with six transport coefficients,~$\zeta,\eta,\kappa,\tau_{\cal E},\tau_{\cal P},\tau_{\cal Q}$. The first three are associated with bulk viscosity, shear viscosity, thermal conductivity, respectively, whereas the latter three are referred to as causal regulators. Here, we have introduced
\begin{align}
    \Theta:=&\nabla_\mu u^\mu,\quad \sigma_{\mu\nu}:=\Delta^\alpha_\mu \Delta^\beta_\nu \left[\nabla_{(\alpha}u_{\beta)}-\frac{1}{3}g_{\alpha\beta}\Theta\right],
\end{align}
which represent the expansion and shear distortion of the fluid element, respectively. In Appendix~\ref{Appendix:BDNK}, we provide the explicit expressions for the nonequilibrium amplitudes for the BDNK fluid.

The BDNK fluid exhibits modal stability in flat space, and is causal, locally well posed, and strongly hyperbolic. Causality is ensured if the following inequalities are satisfied~\cite{Bemfica:2020zjp}:
\begin{align}
    &\tau_{\cal E}, \tau_{\cal Q}, \tau_{\cal P}>0,\quad \eta, \zeta,\kappa>0,\quad \rho \tau_{\cal Q}>\eta,\\
    &\left[\tau_{\cal E} \left(\rho c_s^2 \tau_{\cal Q}+\zeta+\frac{4\eta}{3}+\kappa K_s\right)+\rho \tau_{\cal P} \tau_{\cal Q} \right]^2 \nonumber \\
    &\ge 4 \rho \tau_{\cal E} \tau_{\cal Q} \left[\tau_{\cal P}\left(\rho c_s^2 \tau_{\cal Q}+\kappa K_s\right)-\beta_{\cal E}\left(\zeta+\frac{4\eta}{3}\right)\right]>0,\\
    & 2\rho \tau_{\cal E} \tau_{\cal Q} > \tau_{\cal E} \left(\rho c_s^2 \tau_{\cal Q}+\zeta+\frac{4\eta}{3}+\kappa K_s\right)+\rho \tau_{\cal P}\tau_{\cal Q} >0,\\
    &\rho\tau_{\cal E}\tau_{\cal Q}+\kappa K_s \tau_{\cal P}>\tau_{\cal E}\left(\rho c_s^2 \tau_{\cal Q}+\zeta+\frac{4\eta}{3}+\kappa K_s\right)\nonumber\\
    &+\rho \tau_{\cal P}\tau_{\cal Q}\left(1-c_s^2\right)+\beta_{\cal E}\left(\zeta+\frac{4\eta}{3}\right),
\end{align}
where
\begin{align}
    \rho:=&\varepsilon+p,\\
    \beta_{\cal E}:=&\tau_{\cal Q}\left(\frac{\partial p}{\partial \varepsilon}\right)_n+\frac{\kappa T \left(\varepsilon+p\right)}{n}\left(\frac{\partial \left(\mu/T\right)}{\partial \varepsilon}\right)_n,\\
    K_s:=&\frac{\rho^2 T}{n}\left(\frac{\partial \left(\mu/T\right)}{\partial \varepsilon}\right)_s.
\end{align}
Strong hyperbolicity holds, provided the above causality conditions are satisfied. The initial-value problem is locally well posed if the above causality conditions are satisfied and the transport coefficients are analytic. Then, the system is modally stable in flat space at linear level if the above causality conditions and the following conditions hold~\cite{Bemfica:2020zjp}:
\begin{align}
    &\left(\tau_{\cal E}+\tau_{\cal Q}\right)\left|B\right|\ge \tau_{\cal E} \tau_{\cal Q} D \ge \rho c_s^2\tau_{\cal E}\tau_{\cal Q}\left(\tau_{\cal E}+\tau_{\cal Q}\right),\\
    &\left(\tau_{\cal E}+\tau_{\cal Q}\right)\left|B\right|D+\rho \tau_{\cal E} \tau_{\cal Q}\left(\tau_{\cal E}+\tau_{\cal Q}\right)E\nonumber\\
    &>\tau_{\cal E} \tau_{\cal Q} D^2 +\rho \left(\tau_{\cal E}+\tau_{\cal Q}\right)^2C,\\
    &c_s^2D-E\ge \rho c_s^4\left(\tau_{\cal E}+\tau_{\cal Q}\right),\\
    &\left(\tau_{\cal E}+\tau_{\cal Q}\right)\left[\left|B\right|\left(c_s^2D-2E\right)+2c_s^2\rho \tau_{\cal E}\tau_{\cal Q}E+CD\right]\nonumber\\
    &>2c_s^2\rho \left(\tau_{\cal E}+\tau_{\cal Q}\right)^2 C+\tau_{\cal E}\tau_{\cal Q} D\left(c_s^2D-E\right),\\
    &\left|B\right|\left|D\right|\left[C\left(\tau_{\cal E}+\tau_{\cal Q}\right)+E\tau_{\cal E}\tau_{\cal Q}\right]+2\rho \tau_{\cal E}\tau_{\cal Q}\left(\tau_{\cal E}+\tau_{\cal Q}\right) C E\nonumber\\
    &> \rho C^2 \left(\tau_{\cal E}+\tau_{\cal Q}\right)^2+\tau_{\cal E}\tau_{\cal Q} \left(CD^2+\rho \tau_{\cal E}\tau_{\cal Q} E^2\right) \nonumber\\
    &+B^2 E\left(\tau_{\cal E}+\tau_{\cal Q}\right),
\end{align}
where
\begin{align}
    B:=&-\tau_{\cal E}\left(\rho c_s^2 \tau_{\cal Q}+\zeta+\frac{4\eta}{3}+\kappa K_s\right)-\rho \tau_{\cal P}\tau_{\cal Q},\\
    C:=&\tau_{\cal P}\left(\rho c_s^2 \tau_{\cal Q}+\kappa K_s\right)-\beta_{\cal E}\left(\zeta+\frac{4\eta}{3}\right),\\
    D:=&\rho c_s^2 \left(\tau_{\cal E}+\tau_{\cal Q}\right)+\zeta +\frac{4\eta}{3}+\kappa K_s,\\
    E:=&\kappa T \rho\left[\left(\frac{\partial p}{\partial \varepsilon}\right)_{n}\left(\frac{\partial \left(\mu/T\right)}{\partial n}\right)_\varepsilon-\left(\frac{\partial p}{\partial n}\right)_{\varepsilon}\left(\frac{\partial \left(\mu/T\right)}{\partial \varepsilon}\right)_n\right].
\end{align}

\subsubsection{Small-transport-coefficient expansion}
We work within a small-transport-coefficient expansion. In other words, we introduce a bookkeeping parameter in front of the transport coefficients, and expand the perturbation variables in terms of it. The assumption for the small transport coefficients is commonly adopted in previous work~\cite{HegadeKR:2024agt,Caballero:2025omv} and is supported, at least for $\eta$ and $\zeta$ by current observational constraints~\cite{Ripley:2023lsq,HegadeKR:2024slr}. Indeed, the current constraints~$\zeta~\lesssim 10^{31}~{\rm g~cm^{-1}~s^{-1}}, \eta\lesssim 10^{28}~{\rm g~cm^{-1}~s^{-1}}$ for GW170817 by Ref.~\cite{Ripley:2023lsq} are translated to
\begin{align}
    M\eta \lesssim& 5.12\times 10^{-6}\left(\frac{M}{1.4M_\odot}\right),\\
    M\zeta\lesssim& 5.12\times10^{-3} \left(\frac{M}{1.4M_\odot}\right).
\end{align}

\subsection{Impact on characteristic modes}

\subsubsection{Structural suppression of the causal-regulator corrections}
Equations~\eqref{eq:BDNKtauE} and~\eqref{eq:BDNKtauP} show that the contributions associated with $\tau_{\cal E}$ and $\tau_{\cal P}$ enter through the same combination, implying that they affect the perturbation in a similar manner. As presented in Eqs.~\eqref{eq:S00BDNK},~\eqref{eq:S0BDNK}, and~\eqref{eq:SOmegaBDNK}, the contributions of $\tau_{\cal E}$ and ${\cal \tau}_p$ enter $S_{00}$, $S_0$, and $S_\Omega$, and are proportional to the common factor $\Xi$ defined by
\begin{align}
    \Xi e^{-i \omega t} Y_{\ell m}=\Delta \varepsilon-\left(\varepsilon+p\right)\frac{\Delta n}{n}.\label{eq:TDeltasmuDeltaY}
\end{align}
Likewise, the correction associated with $\tau_{\cal Q}$ in Eq.~\eqref{eq:BDNKkappa_tauQ} contributes to $S_{01}$, $S_{0A}$, and $S_{0A}^-$ in Eqs.~\eqref{eq:S01BDNK},~\eqref{eq:S0ABDNK}, and~\eqref{eq:BDNKS0AOdd}, and it takes the form of the combination of the vector field~${\cal R}^\mu$ defined by
\begin{align}
    {\cal R}^\mu=\Delta^{\mu\nu} \delta \left(\nabla_\sigma T^\sigma_{~\nu}\right),\label{eq:Rmu}
\end{align}
and $\Xi$, as well as the scalar~$\Xi_\alpha$ defined through
\begin{align}
 \Xi_\alpha e^{-i \omega t}Y_{\ell m}=\alpha_1 n T \Delta s+n \mu_i \alpha_{2i}\Delta Y^i,\label{eq:Xialpha}
\end{align}
Here, $\alpha_1$ and $\alpha_{2i}$ are defined below Eq.~\eqref{eq:thermodynamicalrelation}. The components of ${\cal R}^\mu$ are provided in Eqs.~\eqref{eq:vectorR} and~\eqref{eq:vectorROdd}.

Now, Eq.~\eqref{eq:Deltaepsilon} implies that $\Xi$ is of the same order as the out-of-equilibrium correction. Likewise, the second equation in Eq.~\eqref{eq:fieldeqs} indicates that ${\cal R}^\mu$ is at order of the out-of-equilibrium contributions. Moreover, Eq.~\eqref{eq:Xialpha} indicates that $\Xi_\alpha$ is non-vanishing only for non-adiabatic perturbations. In other words, the terms with $\Xi$, $\Xi_\alpha$, and ${\cal R}^\mu$ coupled to the causal regulators are higher order in the small-transport-coefficient expansion on shell. Therefore, the contributions associated with the causal regulators are generically suppressed within the small-transport-coefficient expansion around the perfect-fluid limit. This result indicates that, for small transport coefficients, these contributions have a modest impact on characteristic frequency shifts in the small-transport-coefficient regime.

One might expect that a choice of $(\eta,\zeta,\kappa,\tau_{\cal E},\tau_{\cal  P}, \tau_{\cal Q})$ incompatible with the BDNK conditions mentioned before would introduce immediate pathologies in the characteristic oscillation spectrum. However, our results suggest that this is not necessarily the case, even in the extreme case $\tau_{\cal E}=\tau_{\cal P}=\tau_{\cal Q}=0$. This does not imply inconsistencies with the causality theorem for the BDNK fluid, as the present analysis relies on spherical harmonics and Fourier decompositions, and hence, does not fully capture the principal structure.

\subsubsection{Mode shift}

It is natural to expect that the out-of-equilibrium corrections induce characteristic mode-frequency shifts. We quantify the shifts induced by shear and bulk viscosity within the small-transport-coefficient expansion by assuming zero thermal conductivity, i.e., $\kappa=0$. 
In this work, we parameterize $\eta$ and $\zeta$ as
\begin{align}
    \eta=K_\eta \left(\frac{\varepsilon}{\rho_{\rm sat}}\right)^2,\quad 
    \zeta=K_\zeta \left(\frac{\varepsilon}{\rho_{\rm sat}}\right)^2,\label{eq:Ketazeta}
\end{align}
where $K_\eta$ and $K_\zeta$ are non-negative constants with inverse length scale in the geometric unit and $\rho_{\rm sat}\simeq 2.8\times 10^{14} {\rm~g~cm^{-3}}$ represents the nuclear saturation density. The choice of the exponent of $2$ is phenomenologically motivated. For shear viscosity, the above parametrization can be viewed as phenomenological modeling of the density profile of the electron-electron scattering contribution,~$M \eta \simeq 2.4 \times 10^{-8}(M/1.4M_\odot) (T/10^5{\rm K})^{-2}(\varepsilon/\rho_{\rm sat})^2$, with temperature~$T$~\cite{1976ApJ...206..218F,1987ApJ...314..234C}. For bulk viscosity, the above choice can be interpreted as a phenomenological radial profile motivated by the modified Urca reactions~\cite{Sawyer:1989dp}.

Throughout the analysis, we assume the rest-mass polytrope:
\begin{align}
    \label{eq:poly}\varepsilon=\left(\frac{p}{\kappa_{\rm p}}\right)^{n/(n+1)}+n p,\quad \kappa_{\rm p}= 100 M_\odot^2,\quad n=1.
\end{align}
The regularity condition for the background configuration is given by
\begin{align}
    m=&\frac{4\pi}{3}\varepsilon_c r^3+{\cal O}\left(r^5\right),\\
    \nu=&\nu_c+{\cal O}\left(r^2\right),\quad p=p_c+{\cal O}\left(r^2\right),\nonumber
\end{align}
where $p_c$ is determined through the equation of state for a given central energy density~$\varepsilon_c:=\varepsilon|_{r=0}$. The constant~$\nu_c$ is fixed by imposing the matching condition to the Schwarzschild solution at the stellar surface~$r=R$, i.e., $e^\nu=1-2M/R$. Henceforth, stellar compactness is fixed at $M/R\simeq 0.1460$.

We mainly consider even-parity perturbations. Within the present small-transport-coefficient expansion, the perturbation variables are characterized by two parameters, i.e., $K|_{r=0}$ and $W|_{r=0}$, as in the perfect-fluid case~\cite{Detweiler:1985zz}. Specifically, the regular solutions for the perturbation take the form
\begin{align}
    H_1=&\frac{2}{\ell\left(\ell+1\right)}\left[\ell K_c+8\pi W_c\left(\varepsilon_c+p_c\right)\right]\label{eq:H1exp}\\
    &+\left(C_{H_1 2}^{\rm PF}+C_{H_1 2}^{\eta\zeta}\right)r^2+{\cal O}\left(r^4\right),\nonumber\\
    K=&K_c+\left(C_{K 2}^{\rm PF} + C_{K 2}^{\eta \zeta}\right)r^2+{\cal O}\left(r^4\right),\label{eq:regularK}\\
    W=&W_c+\left(C_{W 2}^{\rm PF} + C_{W 2}^{\eta \zeta}\right)r^2+{\cal O}\left(r^4\right),\label{eq:regularW}\\
    X=&\frac{e^{-\nu_c /2}}{6\ell}\left(\varepsilon_c+p_c\right)\label{eq:regularX}\\
    &\times \left[\ell e^{\nu_c}\left\{3 K_c+ 8\pi W_c\left(\varepsilon_c+3p_c\right)\right\}-6\omega^2 W_c\right]\nonumber\\
    &+i \omega \eta_c\left[\frac{2\varepsilon_c^{(1)} K_c}{3}+W_c\left\{\frac{8\pi}{3\ell}\left(\left(\ell-7\right)\varepsilon_c+3\left(\ell-3\right)p_c\right)\right.\right.\nonumber\\
    &\left.\left.+\frac{4\varepsilon_c^{(1)}}{9}\left(4\pi \left(\varepsilon_c+3p_c\right)-\frac{3 e^{-\nu_c} \omega^2}{\ell}\right)+\frac{2\left(\ell-1\right)}{\ell}\frac{\eta_c^{(2)}}{\eta_c}\right\}\right] \nonumber\\
    &+i\omega \zeta_c \frac{e^{-\nu_c}{\varepsilon_c^{(1)}}}{6\ell} \nonumber\\
    &\times \left[\ell e^{\nu_c}\left\{3 K_c+ 8\pi W_c\left(\varepsilon_c+3p_c\right)\right\}-6\omega^2 W_c\right]\nonumber \\
    &+\left(C_{X 2}^{\rm PF} + C_{X 2}^{\eta \zeta}\right) r^2 +{\cal O}\left(r^4\right),\nonumber
\end{align}
and 
\begin{align}
    H=&K_c-i \omega  \eta_c \frac{32\pi e^{-\nu_c/2}}{\ell}W_c \\
    &+ \left(C_{H2}^{\rm PF}+ C_{H 2}^{\eta \zeta}\right) r^2+{\cal O}\left(r^4\right),\nonumber\\
    H_2=&K_c + \left(C_{H 2}^{\rm PF}+ C_{H_2 2}^{\eta \zeta}\right)r^2+{\cal O}\left(r^4\right),\\
    V=&-\frac{W_c}{\ell}+\left(C_{V2}^{\rm PF}+ C_{V2}^{\eta \zeta}\right) r^2+{\cal O}\left(r^4\right),\label{eq:Vexp}
\end{align}
where $\varepsilon_c^{(1)}= d\varepsilon/dp|_{r=0}$, $\eta_c= \eta|_{r=0}$, $\eta_c^{(2)}=d^2\eta/dr^2|_{r=0}$. The coefficients~$C_{q2}^{\rm PF}$ with $q=H_1,K,W,X,H,V$ denote the second-order Taylor coefficients in the perfect-fluid sector and are fixed by the two central amplitudes~$K_c$ and $W_c$ for a given stellar model. The coefficients~$C_{q2}^{\eta \zeta}$ with $q=H_1,K,W,X,H,H_2,V$ denote the BDNK linear corrections and vanish in the limit~$\eta,\zeta\to0$. Then, imposing $\Delta p=0$ at the stellar surface fixes the ratio~$W_c/K_c$, while the remaining parameter merely rescales the perturbation variables and can therefore be set to unity without loss of generality. This surface condition is commonly imposed for the perfect-fluid case, where the normal and tangential components of traction at the surface are required to vanish. We expect this to be sufficient for the current viscous-fluid configuration considered in our perturbative numerical analysis. This is because, for the particular parameterization of the radial profiles of the transport coefficients in Eq.~\eqref{eq:Ketazeta}, the out-of-equilibrium contributions to the surface traction are proportional to $\varepsilon^2$ and hence vanish at the surface. Consequently, the required surface conditions reduce to the perfect-fluid condition~$\Delta p=0$ at the surface. In Section~\ref{sec:newfamily}, we discuss appropriate surface conditions in the non-perturbative analysis with generic radial profiles of transport coefficients.

\begin{figure*}[t]
\centering
 \includegraphics[scale=0.5]{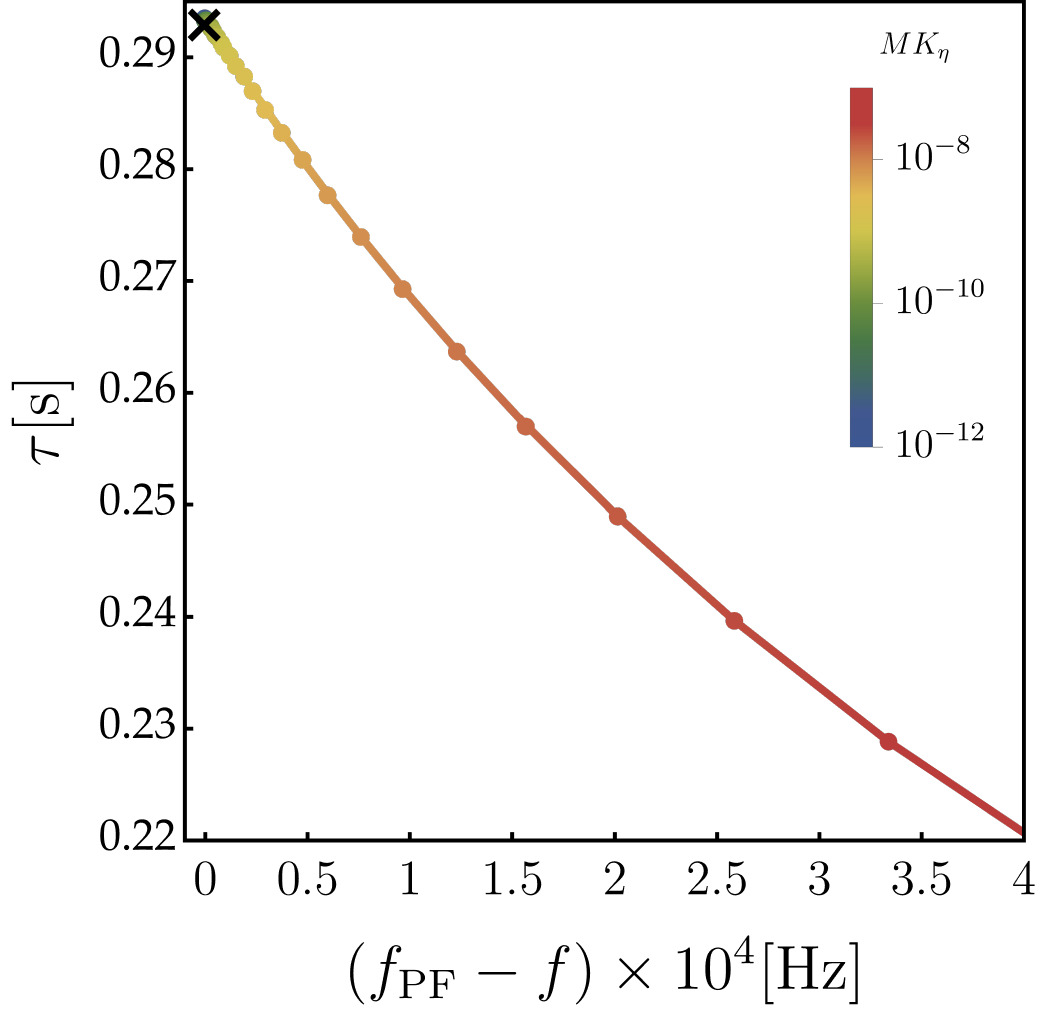}
  \includegraphics[scale=0.5]{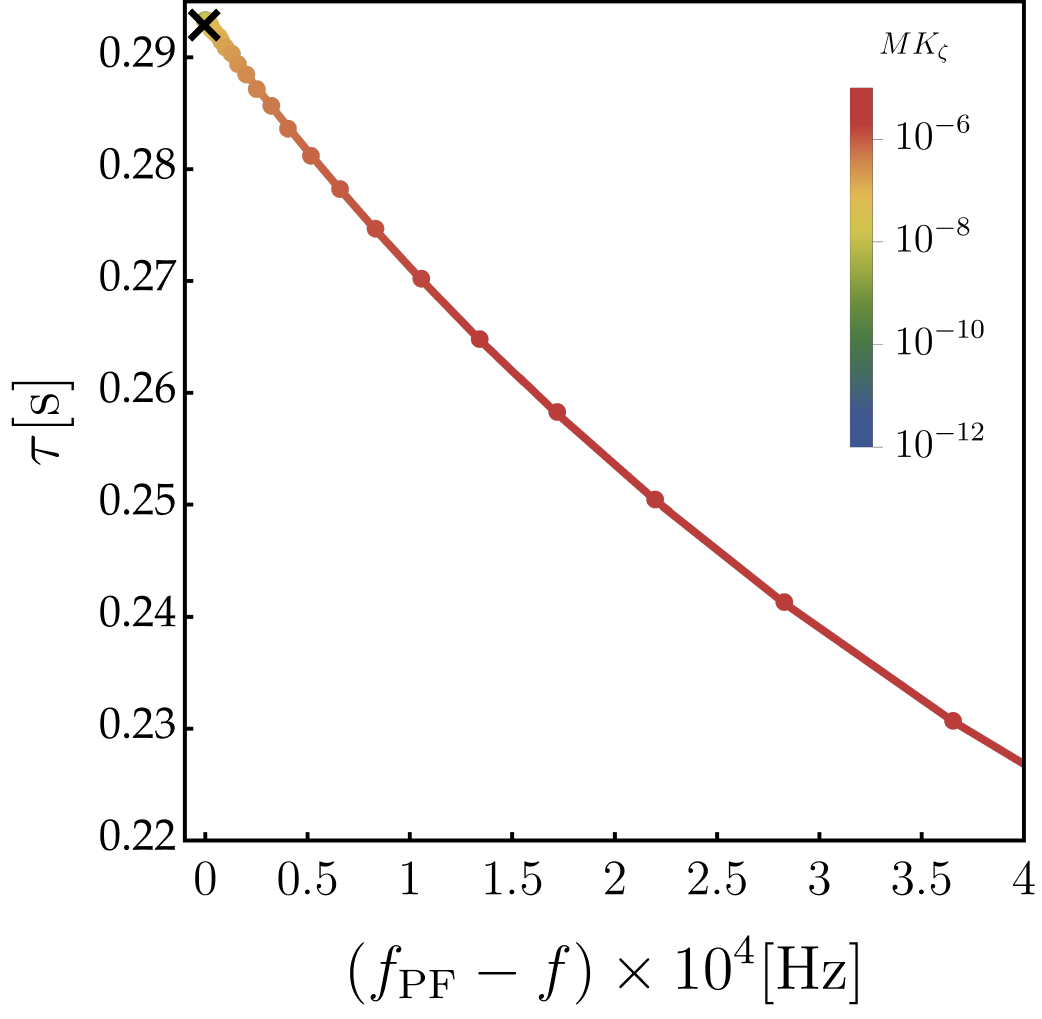}
\caption{The quadrupolar~$(\ell=2)$~$f$-mode shifts induced by shear viscosity~(left panel) and bulk viscosity~(right panel) with the phenomenological coefficients in Eq.~\eqref{eq:Ketazeta} and the rest-mass polytropic equation of state in Eq.~\eqref{eq:poly} for a fixed stellar compactness of $M/R \simeq 0.1460$. The black cross denotes the $f$-mode frequency and the damping time of the perfect-fluid case,~$(f_{\rm PF},\tau_{\rm PF})\simeq (1.578 \times 10^3{\rm Hz},2.934 \times 10^{-1} {\rm s})$.} 
\label{fig:fmodeshift}
\end{figure*}

\begin{figure*}[t]
\centering
 \includegraphics[scale=1.24]{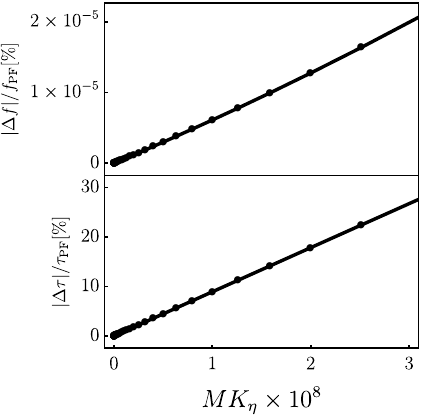}
  \includegraphics[scale=1.24]{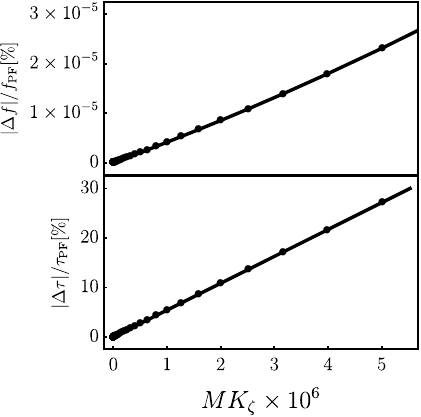}
\caption{Absolute values of the relative shifts of the $f$-modes with respect to the perfect-fluid case. Here, we define $\Delta f =f-f_{\rm PF}$ and $\Delta 
\tau=\tau-\tau_{\rm PF}$.}
\label{fig:Err_f}
\end{figure*}

The regular interior perturbation is matched to the exterior problem through the logarithmic derivative of the Zerilli function, constructed following Ref.~\cite{Sotani:2001bb}. In the exterior, we map the even-parity problem to the Regge-Wheeler equation via the Chandrashekar transformation~\cite{1983mtbh.book.....C}. The characteristic mode frequencies are then determined by imposing the outgoing-wave condition at infinity and are computed using the continued fraction method outlined in Ref.~\cite{Sotani:2001bb}. To characterize stellar oscillations, we introduce the oscillation frequency~$f$ and the damping timescale $\tau$ through
\begin{align}
    \omega = 2\pi f - \frac{i}{\tau}.
\end{align}
Note that, with our convention~$e^{-i \omega t}$ for the Fourier decomposition, $f={\rm Re}[\omega]/(2\pi)$, while $\tau=-1/{\rm Im[\omega]}$. In what follows, we analyze the fundamental fluid~($f$-) mode~\cite{Cowling:1941nqk} and spacetime~($w$-) mode~\cite{1992MNRAS.255..119K}.

Figure~\ref{fig:fmodeshift} presents the $f$-mode shifts induced by shear viscosity~(left panel) and bulk viscosity~(right panel) within the small-transport-coefficient expansion.  For the particular parametrization of the transport coefficients considered, we find that the viscous effects tend to decrease both $f$ and $\tau$. The trend in $\tau$ can be directly interpreted as a consequence of viscosity providing an additional dissipative channel for the mode, thereby increasing the damping rate.

\begin{figure*}[t]
\centering
 \includegraphics[scale=0.50]{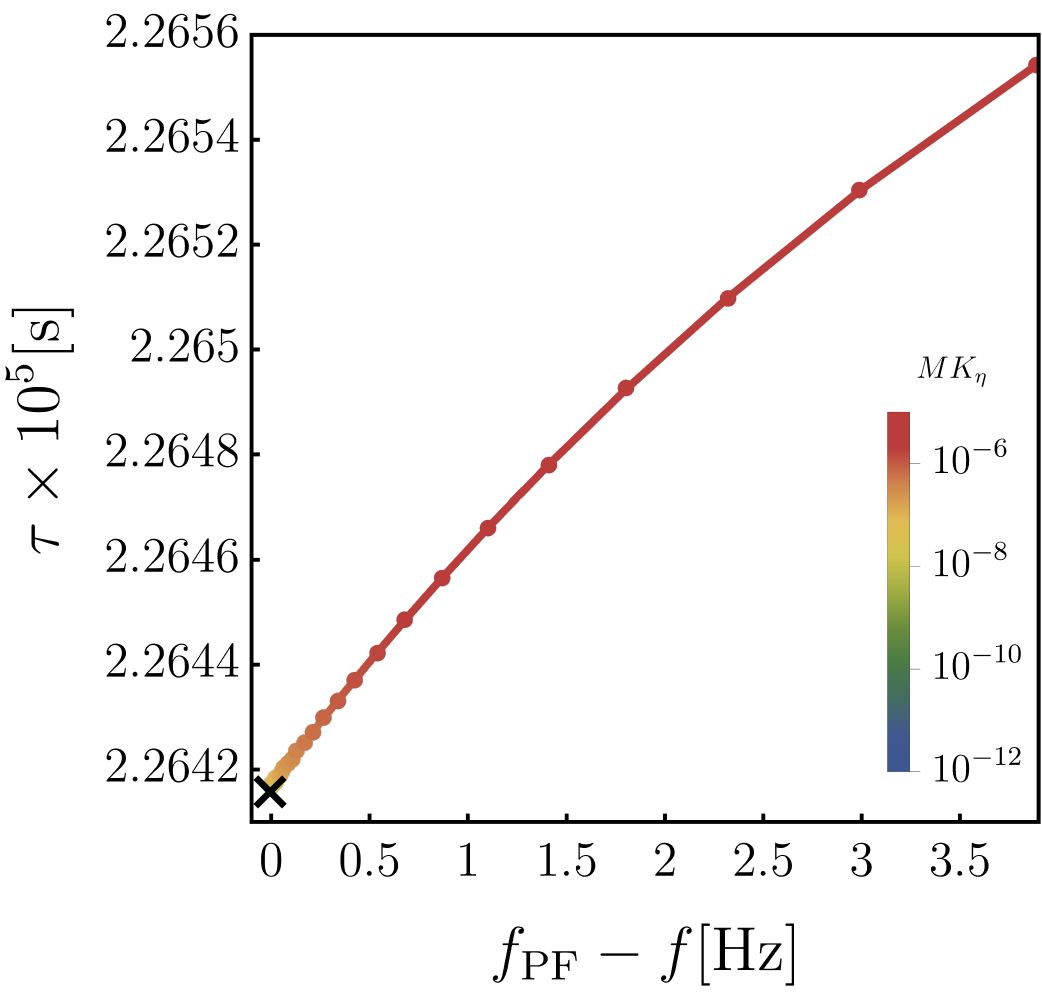}
  \includegraphics[scale=0.505]{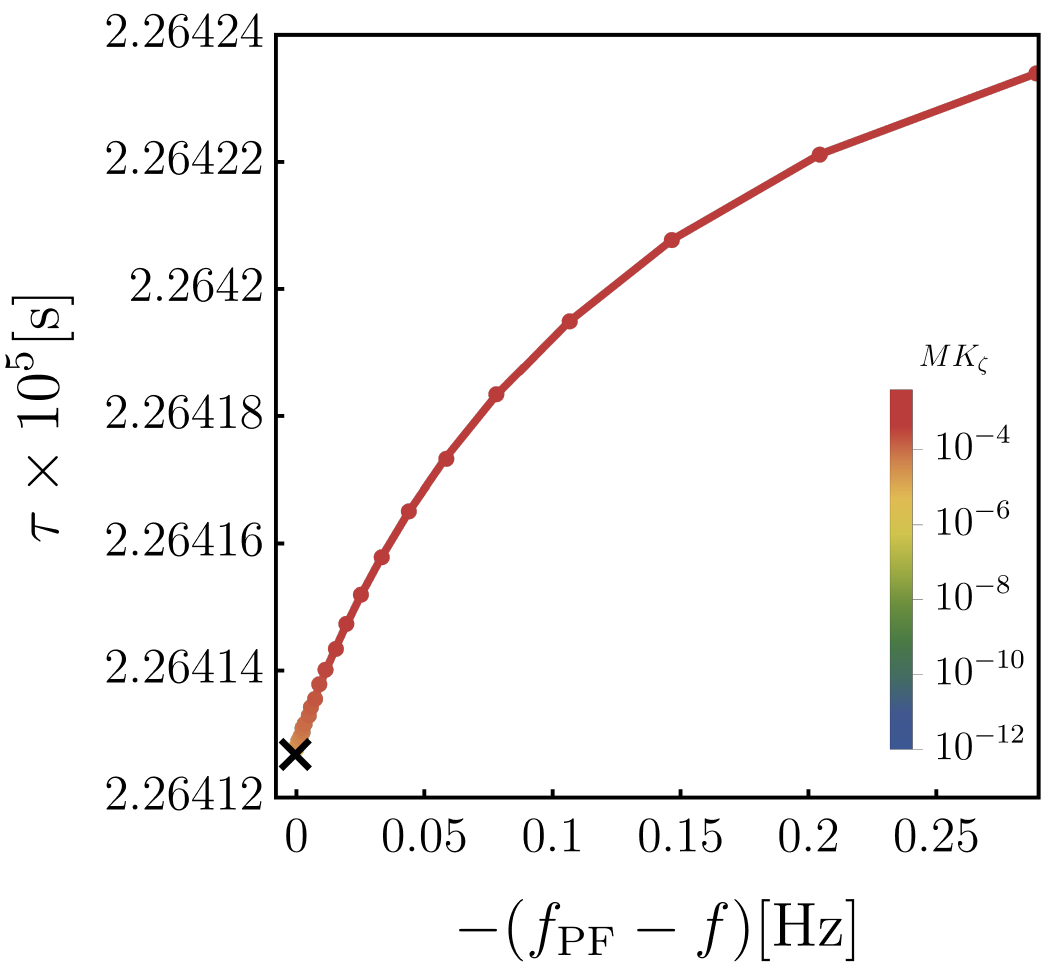}
\caption{The quadrupolar~$(\ell=2)$~even-parity $w$-mode shifts induced by shear viscosity~(left panel) and bulk viscosity~(right panel). Note the opposite sign of the horizontal axis of the right panel. The black cross denotes the $w$-mode frequency and the damping time for the perfect-fluid case,~$(f_{\rm PF},\tau_{\rm PF})\simeq (1.009\times 10^{4}{\rm Hz},2.264\times 10^{-5}~{\rm s})$. The equation of state and stellar compactness are the same as in Fig.~\ref{fig:fmodeshift}.} 
\label{fig:wmodeshift}
\end{figure*}

\begin{figure*}[t]
\centering
 \includegraphics[scale=1.24]{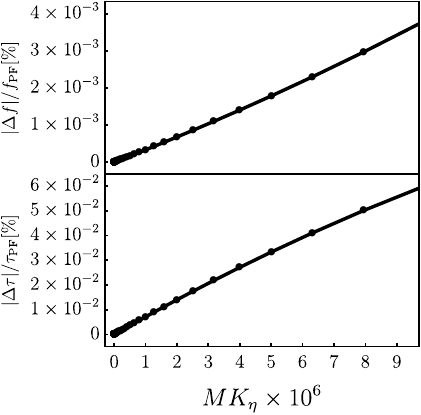}
  \includegraphics[scale=1.24]{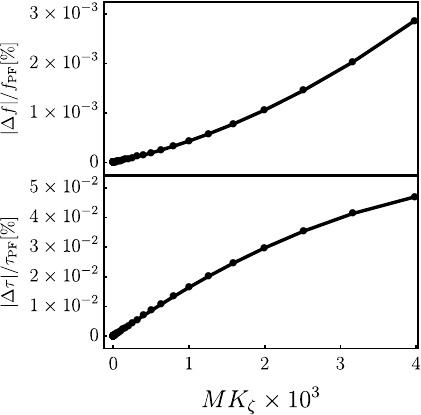}
\caption{Absolute values of the relative shifts of the $w$-modes with respect to the perfect-fluid case. Here, we define $\Delta f =f-f_{\rm PF}$ and $\Delta 
\tau=\tau-\tau_{\rm PF}$.}
\label{fig:Err_w}
\end{figure*}

\begin{figure*}[t]
\centering
 \includegraphics[scale=0.525]{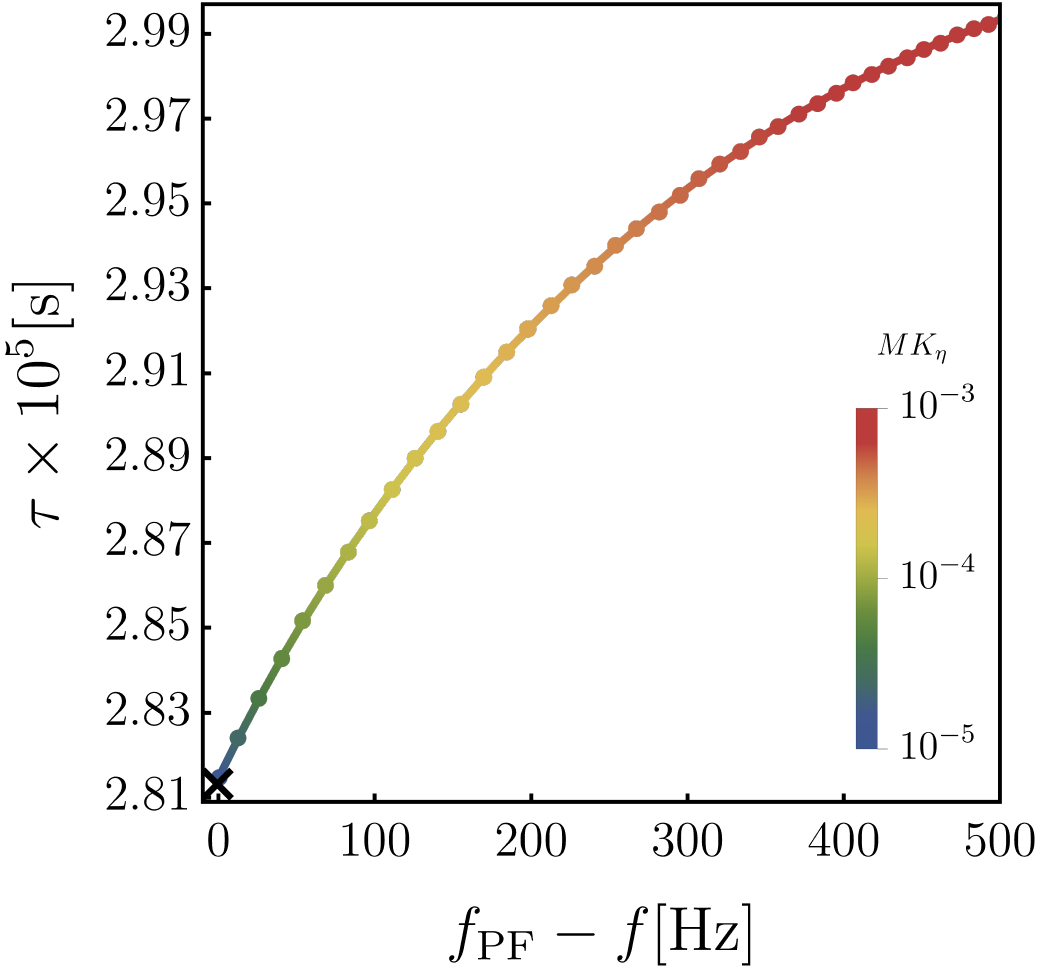}
  \includegraphics[scale=1.24]{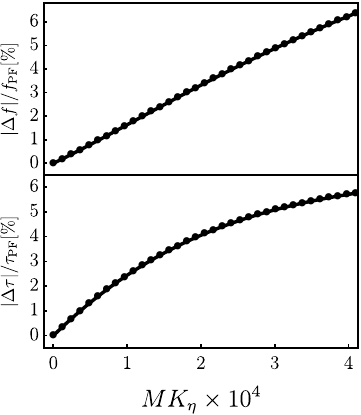}
\caption{The quadrupolar~$(\ell=2)$ odd-parity $w$-mode shifts~(left panel) and the corresponding absolute values of the relative shifts with respect to the perfect-fluid case~(right panel). Here, we define $\Delta f =f-f_{\rm PF}$ and $\Delta 
\tau=\tau-\tau_{\rm PF}$. The black cross denotes the odd-parity $w$-mode frequency and the damping time for the perfect-fluid case,~$(f_{\rm PF}, \tau_{\rm PF}) \simeq (7.088\times 10^3{\rm Hz},2.815\times 10^{-5}{\rm s})$. The equation of state and stellar compactness are the same as in Fig.~\ref{fig:fmodeshift}.}
\label{fig:modeshift_odd}
\end{figure*}

Figure~\ref{fig:Err_f} shows the absolute values of the relative shifts of the $f$-mode with respect to the perfect-fluid case. Their approximately linear dependence on the transport coefficients indicates that the small-transport-coefficient expansion is self-consistent within this regime. It is useful to compare our results with the estimates based on the approximate formulae for the damping rates induced by shear and bulk viscosity,
$
    {1}/{\tau_\eta}={5\eta}/({\varepsilon R^2})$ and $
    {1}/{\tau_\zeta}=({324}/{4375}){\zeta}/({\varepsilon R^2})
$~\cite{1987ApJ...314..234C}.
Assuming $\varepsilon\sim 3M/(4\pi R^3)$, these estimates give the corresponding shifts
\begin{align}
\frac{\Delta \tau_\eta}{\tau_{\rm PF}}=-\frac{20\pi R}{3M}\eta \tau_{\rm PF},\quad \frac{\Delta \tau_\zeta}{\tau_{\rm PF}}=-\frac{432\pi R}{4375M}\zeta \tau_{\rm PF},\label{eq:tauestimate}
\end{align}
to linear order in the transport coefficients. For the current setup, these estimates give approximately $|\Delta \tau_\eta|/\tau_{\rm PF}\simeq 6$\% for $\eta M=10^{-8}$ and $|\Delta \tau_\zeta|/\tau_{\rm PF} \simeq 9$\% for $\zeta M=10^{-6}$, respectively, in agreement with our numerical results within the perturbative treatment based on the small-transport-coefficient expansion around the perfect-fluid limit. Moreover, for the transport coefficients of the same order of magnitude, shear viscosity induces a larger shift than bulk viscosity by nearly two orders of magnitude. This is also consistent with the estimate of Ref.~\cite{1987ApJ...314..234C}, as can be seen directly from Eq.~\eqref{eq:tauestimate}. The absolute values of the relative shift of $\tau$ can reach $10\%$ at $(MK_\eta,MK_\zeta)=({\cal O}(10^{-8}),{\cal O}(10^{-6}))$, beyond which the present perturbative treatment becomes no longer a good approximation.

Now, we turn to the $w$-mode. Figure~\ref{fig:wmodeshift} shows the shifts of the even-parity $w$-mode induced by shear viscosity~(left panel) and bulk viscosity~(right panel) within the small-transport-coefficient expansion. The shear-viscosity-induced shift in $f$ is negative, whereas the bulk-viscosity-induced shift is positive. A common intriguing feature of the $w$-mode shifts found here is that $\tau$ increases as the viscous effects become stronger, which is not observed in the $f$-mode shifts. In other words, in such a case, viscosity reduces the damping rate of the $w$-modes. A qualitatively similar trend was found in the axial perturbations of BDNK viscous stars~\cite{Boyanov:2024jge,Bussieres:2026rnz}. Our results show that (i) the even-parity $w$-modes can exhibit the same trend, and (ii) bulk viscosity gives rise to the same effects.

Figure~\ref{fig:Err_w} presents the absolute values of the relative shifts of the $w$-mode. Remarkably, they remain less than $0.1\%$ even at larger transport coefficients than those in the $f$-mode case. Our results are typically restricted up to ${\cal O}(10^{-2})$ for the magnitude of $MK_\eta$ and $MK_\zeta$, beyond which the values found are unreliable. As in the $f$-mode, for the transport coefficients of the same order of magnitude, shear viscosity
induces a larger shift than bulk viscosity, but by nearly three orders of magnitude.

Now, we briefly compare our results with previous studies of {\it radial} oscillations of BDNK stars. First, our results imply that viscous effects have a stronger impact on the damping rate than on the oscillation frequency. This trend has also been observed in radial oscillations~\cite{Shum:2025jnl,Keeble:2026bzo}. Another intriguing finding in both the $f$- and $w$-modes is that, within our particular parametrization of the radial profiles of transport coefficients as given in Eq.~\eqref{eq:Ketazeta}, shear viscosity induces a larger shift than bulk viscosity, for the transport coefficients of the same order of magnitude. This result is notably different from the observation in radial oscillations. In such cases, the expansion of fluid elements is expected to dominate over shear distortion, and hence, the bulk-viscous effects may play the leading role in mode shifts~\cite{Shum:2025jnl}. By contrast, in non-radial oscillations, shear distortion may be more efficiently induced because the perturbation is no longer spherically symmetric. This may provide a qualitative explanation for why shear viscosity has a stronger impact in our results.

Finally, Fig.~\ref{fig:modeshift_odd} presents the odd-parity $w$-mode shifts induced by shear viscosity, for comparison with the even-parity results shown in the left panels of Figs.~\ref{fig:wmodeshift} and~\ref{fig:Err_w}. As shown in the left panel of Fig.~\ref{fig:modeshift_odd}, the oscillation frequency~$f$ decreases, whereas the damping timescale~$\tau$ increases. These features are qualitatively consistent with the trends observed in the even-parity sector. The right panel shows the corresponding relative shifts. For both $f$ and $\tau$, the absolute values of the relative shifts reach $1\%$ at $M K_\eta={\cal  O}(10^{-4})$, beyond which the data no longer exhibit an approximately linear dependence on $M K_\eta$.

Why does the damping time for the $w$-modes increase (i.e., exhibit less dissipation) as viscosity is increased? We now briefly discuss this curious behavior directly in terms of the odd-parity perturbation equation, 
obtained from Eq.~\eqref{eq:RWinterior} with the BDNK nonequilibrium amplitudes given in Eqs.~\eqref{eq:BDNKS1AOdd} and~\eqref{eq:BDNKSZOdd}, 
    \begin{align}
    &e^{\left(\nu-\lambda\right)/2}\left(e^{\left(\nu-\lambda\right)/2} \psi'\right)'\nonumber\\
    &+\left[\omega^2-e^\nu \left(\frac{\ell\left(\ell+1\right)}{r^2}-\frac{6m}{r^3}+4\pi \left(\varepsilon-p\right)\right)\right]\psi\label{eq:eqforpsiinBDNK}\\
    &=-16 \pi e^{\nu/2}i \omega \eta \psi-\frac{4e^{\left(3\nu-\lambda\right)/2}}{r\left(\varepsilon+p\right)}\left(e^{\nu/2}\eta\right)'U.\nonumber
\end{align}
Note that this takes the form of a forced damped harmonic oscillator. If one instead considers the homogeneous equation for $\psi$, obtained by setting $U=0$ in Eq.~\eqref{eq:eqforpsiinBDNK}, the equation reduces to that of a damped harmonic oscillator, for which we find that the damping time decreases with increasing viscosity, as expected. This indicates that the unusual damping behavior of the axial $w$-modes originates from the coupling between $\psi$ and $U$. It would be interesting to investigate whether the behavior of the even-parity $w$-mode can be understood in a similar way.

\begin{figure}[t]
\centering
 \includegraphics[scale=0.47]{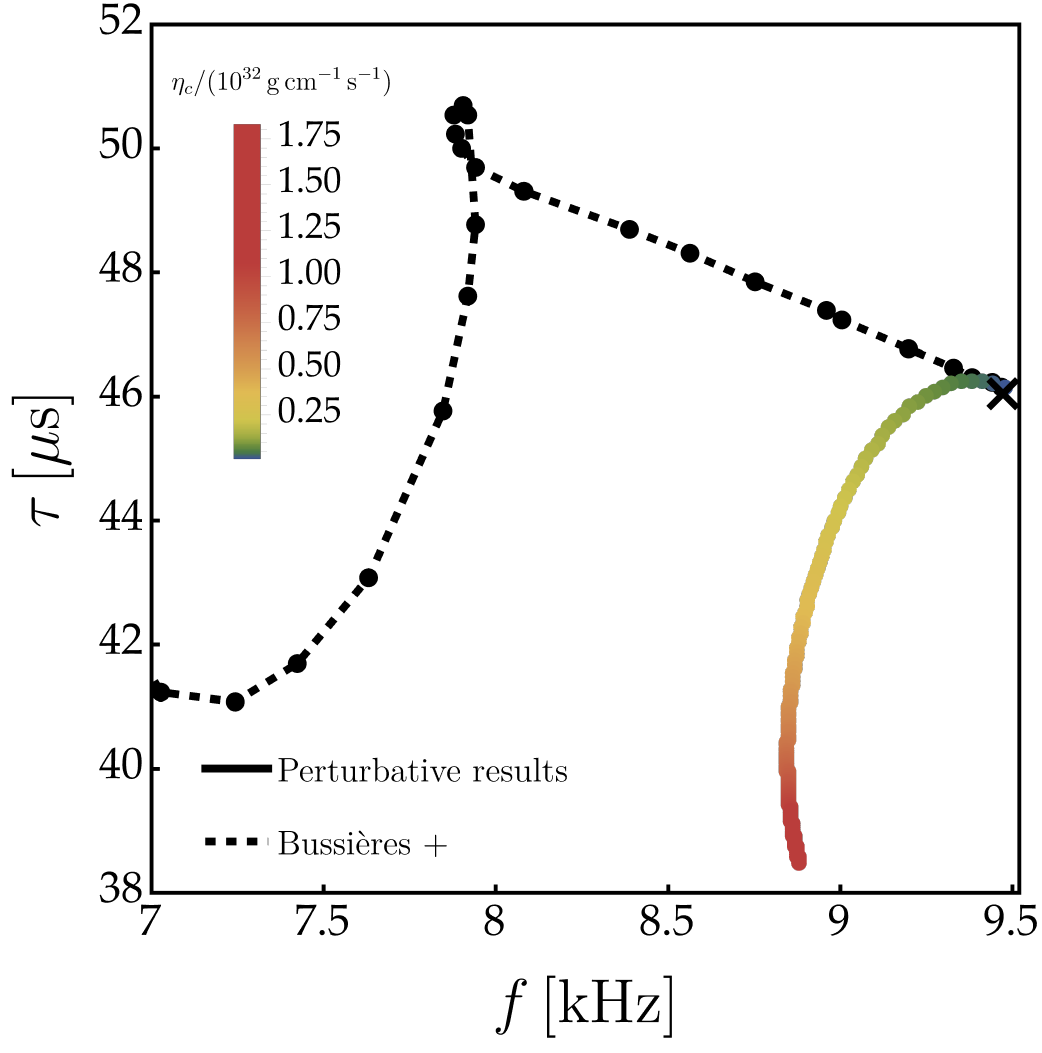}
\caption{Comparison of our perturbative results with the non-perturbative results for the ``A2'' model in Bussi\`{e}res et al.~\cite{Bussieres:2026rnz}. The notation follows the reference.}
\label{fig:Comparison}
\end{figure}

\subsection{Towards non-perturbative regimes: New family}\label{sec:newfamily}

Although the analysis in the previous section relies on a perturbative expansion of the perturbation variables in terms of the transport coefficients, our results still provide useful insight into the mode structure beyond a perturbative treatment. In the small-transport-coefficient regime, the causal regulators are expected to have only a modest effect on the mode shifts shown in the previous section, due to their structural suppression, provided that their magnitudes are not significantly larger than shear and bulk viscous coefficients. This expectation is supported, at least for the odd-parity sector, by a direct comparison with the non-perturbative results presented in Fig.~2 of Ref.~\cite{Bussieres:2026rnz} within the same parameter setting, as shown in Fig.~\ref{fig:Comparison}.

As discussed in Sections~\ref{Section:EvenParity} and~\ref{Section:OddParity}, the BDNK corrections raise the differential order of the perturbation equations relative to the perfect-fluid case, potentially leading to additional mode branches. Such branches have no regular perfect-fluid counterparts in the limit of vanishing transport coefficients. Therefore, the present (regular) perturbative analysis does not capture them. Indeed, within our perturbative system, we do not find the axial ``$\eta$-mode'' of BDNK viscous stars, a new mode family discovered in the non-perturbative analysis of Ref.~\cite{Bussieres:2026rnz}, even when the mode frequencies reported therein are used as initial guesses in our root search. It is worth noting that the BDNK fluid model is a first-order hydrodynamic effective theory based on the gradient expansion about zeroth-order perfect-fluid constitutive relations. Therefore, these new branches should be interpreted as features of the closed eigenvalue problem, rather than robust controlled predictions within the effective-theory framework. 

In what follows, we discuss further how new families of modes may emerge in the non-perturbative system by analyzing the Taylor expansions of the perturbations around $r=0$.

\subsubsection{Odd parity}
First, we focus on the $\ell=2$ odd-parity perturbation. Let us first consider the perfect-fluid case. It follows from the corresponding counterpart of Eq.~\eqref{eq:RWinterior} that the regular solution of $\psi$ at $r=0$ takes the form
\begin{align}
    \psi=&\psi_c r^3+C_{\psi5}^{\rm PF}  r^5+C_{\psi7}^{\rm PF} r^7+{\cal O}\left(r^9\right),\label{eq:PFpsi}
\end{align}
where $\psi_c$ is an arbitrary constant. The coefficients are given by
\begin{align}
    C_{\psi5}^{\rm PF}=&\frac{\psi_c}{14}\left[16\pi \left(\varepsilon_c-p_c\right)-e^{-\nu_c}\omega^2\right],\label{eq:C5PF}\\
    C_{\psi7}^{\rm PF}=&-\frac{e^{-2\nu_c}\psi_c}{1512}\left[64 e^{2\nu_c}\pi^2\left\{3 p_c^2\left(7 \varepsilon_c^{(1)}-13\right)\right.\right.\nonumber\\
    &\left.+4 \left(7\varepsilon_c^{(1)}+8\right)p_c \varepsilon_c+7\left(\varepsilon_c^{(1)}-7\right)\varepsilon_c^2\right\} \label{C7PF}\\
    &\left.+32 \pi e^{\nu_c}\left(5\varepsilon_c-9p_c\right)\omega^2-3\omega^4 \right].\nonumber
\end{align}
It is worth noting that, in the perfect-fluid case, the regular solution for $\psi$ contains only the overall normalization scale~$\psi_c$, owing to decoupling from $U$. Then, $U$ is determined algebraically and is expanded as
\begin{align}
    U=C_{U0}^{\rm PF} r^3+{\cal O}\left(r^5\right),
\end{align}
with 
\begin{align}
    C_{U0}^{\rm PF}=\frac{16\pi}{i \omega}e^{-\nu_c/2}\left(\varepsilon_c+p_c\right)\psi_c.\label{eq:CU0PF}
\end{align}

By contrast, in the BDNK-fluid case, $U$ satisfies a second-order differential equation in Eq.~\eqref{eq:EqforUinpsi} and is coupled to $\psi$. We now assume that the odd-order derivatives of $\eta$ and $\tau_Q$ vanish at $r=0$, as expected from the smooth background quantities near the center. The regular solution of $\psi$ then takes the form
\begin{align}
    \psi=&\psi_c r^3+C_{\psi5}^{\rm BDNK}  r^5+C_{\psi7}^{\rm BDNK} r^7+{\cal O}\left(r^9\right),\label{eq:viscouspsi}
\end{align}
where
\begin{align}
    C_{\psi5}^{\rm BDNK}=&C_{\psi5}^{\rm PF}-\frac{2e^{\nu_c}U_c}{7\left(\varepsilon_c+p_c\right)}\left[\frac{4\pi \left(\varepsilon_c+3p_c\right)}{3}\eta_c+\eta_c^{(2)}\right]\nonumber\\
    &-i \omega \eta_c \frac{8\pi e^{-\nu_c/2}}{7}\psi_c,
\end{align}
and 
\begin{widetext}
\begin{align}
    C_{\psi7}^{\rm BDNK}=&C_{\psi7}^{\rm PF}+\frac{2\pi i\omega}{189}e^{\nu_c/2}\left(\varepsilon_c+3p_c\right) \left[U_c-\frac{16\pi}{i\omega}e^{-\nu_c/2}\left(\varepsilon_c+p_c\right)\psi_c\right]\nonumber\\
&+\psi_c \Bigg[
 \frac{4\pi i\omega}{189}
 e^{-3\nu_c/2}\eta_c
 \left\{
 2e^{\nu_c}\pi(33p_c-53\varepsilon_c)
 +3\omega^2
 \right\} -\frac{32\pi^2}{63}
 e^{-\nu_c}\eta_c^2\omega^2 -\frac{26\pi }{63}
 i \omega e^{-\nu_c/2} \eta_c^{(2)}\nonumber\\
 & -\frac{8\pi}{63}
 (\varepsilon_c+p_c)
 \left(
 1-i\omega e^{-\nu_c/2}\tau_{Qc}
 \right)
 \frac{\eta_c^{(2)}}{\eta_c}+\frac{32\pi^2}{189}
 i\omega e^{-\nu_c/2}
 (\varepsilon_c+p_c)(\varepsilon_c+3p_c)
 \tau_{Qc}
\Bigg]
\\\nonumber
&
+U_c \Bigg[
 \frac{
 32 i\omega e^{\nu_c/2}\pi^2
 (\varepsilon_c+3p_c)\eta_c^2
 }{
 189(\varepsilon_c+p_c)
 } +\frac{2\pi}{189}
 \tau_{Qc}\omega
 \left\{
 12 i e^{\nu_c/2}\eta_c^{(2)}
 +\left(\varepsilon_c+3p_c\right)\omega
 \right\}
\\\nonumber
&
 -\frac{
 e^{\nu_c}
 \left\{
 7\eta_c^{(4)}
 +16\pi \eta_c^{(2)}\left(14\varepsilon_c+3p_c\right)
 \right\}
 -3\eta_c^{(2)}\omega^2
 }{
 378(\varepsilon_c+p_c)
 }+\left\{
 \frac{i\omega e^{\nu_c/2}}{126}
 +\frac{\tau_{Qc}\omega^2}{126}
 +\frac{e^{\nu_c}\eta_c^{(2)}}
 {126(\varepsilon_c+p_c)}
 \right\}
 \frac{\eta_c^{(2)}}{\eta_c}
\\\nonumber
&
 +\eta_c \Bigg[
 \frac{32\pi^2}{189}
 i\omega e^{\nu_c/2}
 (\varepsilon_c+3p_c)\tau_{Qc}+\frac{2\pi}{945(\varepsilon_c+p_c)}
 \Bigg\{
 4e^{\nu_c}\pi(\varepsilon_c+3p_c)
 \left(
 p_c(45+7\varepsilon'_c)
 -(95-7\varepsilon'_c)\varepsilon_c
 \right)\nonumber
\\
&
 +60i \omega e^{\nu_c/2}\eta_c^{(2)}+5(\varepsilon_c+3p_c)\omega^2\Bigg\}\Bigg]\Bigg],\nonumber
\end{align}
\end{widetext}
with $\tau_{Qc}:=\tau_Q|_{r=0}$ and $\eta_c^{(4)}=d^4\eta/dr^4|_{r=0}$. Here, the additional parameter~$U_c$ is unconstrained and is defined by
\begin{align}
U=U_c r^3+{\cal O}(r^5),
\end{align}
for the BDNK fluid. The presence of $U_c$ is a notable difference from the perfect-fluid case and is interpreted as a consequence of the fluid-velocity perturbations becoming a dynamical degree in oscillations. The coefficient~$C_{\psi5}^{\rm BDNK}$ coincides with $C_{\psi5}^{\rm PF}$ when $\eta_c, \eta_c^{(2)}\to0$. However, $C_{\psi7}^{\rm BDNK}$ does not reduce to $C_{\psi7}^{\rm PF}$ in general: assuming that $\eta''$ and $\eta''''$ approach zero sufficiently faster than $\eta$ as $r\to 0$, one can take the smooth limit~$\eta, \tau_Q\to0$, thereby obtaining
\begin{align}
   \lim_{\eta, \tau_Q\to 0} C_{\psi7}^{\rm BDNK} =&C_{\psi7}^{\rm PF}+\frac{2\pi i\omega}{189}e^{\nu_c/2}\left(\varepsilon_c+3p_c\right)\label{eq:PFlimit}\\
&\times \left[U_c-\frac{16\pi}{i\omega}e^{-\nu_c/2}\left(\varepsilon_c+p_c\right)\psi_c\right].\nonumber
\end{align}
It follows that $\psi$ for the BDNK fluid does not, in general, reduce to its perfect-fluid counterpart. This suggests that $U_c$ is expected to be responsible for a new family that does not approach any perfect-fluid modes as the transport coefficients tend to be zero. Equation~\eqref{eq:PFlimit} also shows that, if $U_c$ coincides with $C_{U0}^{\rm PF}$ in Eq.~\eqref{eq:CU0PF}, then $C_{\psi7}^{\rm BDNK}$ reduces to $C_{\psi7}^{\rm PF}$ in the limit of vanishing transport coefficients. This is precisely what occurs in the small-transport-coefficient expansion, where $U$ appearing in the equation for $\psi$ in Eq.~\eqref{eq:eqforpsiinBDNK} is evaluated at the perfect-fluid order and hence, the perturbative treatment selects the perfect-fluid-connected modes.

In the non-perturbative analysis of Ref.~\cite{Bussieres:2026rnz}, the regularity condition for $U$ and $\psi$ at $r=0$, the outgoing-wave condition at infinity for $\psi$, as well as the regularity condition for $U$ and $\psi$ at the stellar surface select $U_c/\psi_c$ and discrete values of $\omega$ simultaneously. This fact further illuminates the underlying structure of characteristic oscillations of BDNK viscous stars. In the small-transport-coefficient regime, if $U_c/\psi_c$ approaches the perfect-fluid value~$C_{U0}^{\rm PF}/\psi_c$, the corresponding eigenvalue is expected to approach a perfect-fluid~$w$-mode frequency. By contrast, if $U_c/\psi_c$ does not approach $C_{U0}^{\rm PF}/\psi_c$, the eigenfunction does not reduce to the perfect-fluid counterpart continuously and is interpreted as a new branch driven by viscosity.

\subsubsection{Even parity}
We now turn to the even-parity sector. We first focus on the shear and bulk viscous contributions in the BDNK fluid. The relevant nonequilibrium amplitudes~$S_0, S_1, S_Z, S_\Omega$, given in Eqs.~\eqref{eq:S0BDNK}--\eqref{eq:SOmegaBDNK}, contain $H_2,K, V, W, V', W'$. Since Eqs.~\eqref{eq:eqforpX} and~\eqref{eq:eqforX} involve derivatives of these amplitudes, i.e., $S_0'$ and $S_1'$, the resulting perturbation equations contain higher-order derivatives such as $W'', H_2', V''$. Thus, the even-parity BDNK system has a higher-derivative structure relative to the perfect-fluid case.

As in the odd-parity sector, we expand the perturbation variables around $r=0$ for the $\ell=2$ mode, and discuss separately the contributions of bulk viscosity~$\zeta$ and shear viscosity~$\eta$. In the perfect-fluid case, there are two unconstrained parameters~$K_c$ and $W_c$ defined in Eqs.~\eqref{eq:regularK} and~\eqref{eq:regularW}. When bulk viscosity is the only out-of-equilibrium contribution, we find that the regular expansion of the solutions is characterized by $K_c$ and $W_c$ to sixth order, as in the perfect-fluid case, and that, in the limit~$\zeta\to 0$, all the Taylor coefficients reduce to their perfect-fluid counterparts given in Eqs.~\eqref{eq:H1exp}--\eqref{eq:Vexp}. Thus, the bulk-viscous contribution does not give rise to additional unconstrained central amplitudes.

By contrast, in the presence of shear viscosity alone, we find one additional unconstrained parameter~$W_c^{(2)}$ defined through
\begin{align}
    W=&W_c+ W_c^{(2)} r^2 +{\cal O}\left(r^4\right).\label{eq:WBDNK}
\end{align}
We have verified that all the Taylor coefficients up to sixth order are then fixed by $W_c,K_c$, and $W_c^{(2)}$. In the limit~$\eta\to 0$, the lowest-order Taylor coefficients of the perturbation variables reduce to their perfect-fluid counterparts given in Eqs.~\eqref{eq:H1exp}--\eqref{eq:Vexp}, whereas the second-order coefficients, denoted by $C_{q2}^{\rm BDNK}$ with $q=H,H_2,K,H_1,V,X$, lead to
\begin{align}
&\lim_{\eta \to0}
\begin{pmatrix}
C_{H2}^{\rm BDNK} \\
C_{H_2 2}^{\rm BDNK} \\
C_{K2}^{\rm BDNK}
\end{pmatrix}
=
\begin{pmatrix}
C_{H2}^{\rm PF} \\
C_{H_2 2}^{\rm PF} \\
C_{K2}^{\rm PF}
\end{pmatrix},
\end{align}
and
\begin{align}
   &\lim_{\eta \to0}C_{H_1 2}^{\rm BDNK}=C_{H_1 2}^{\rm PF}+\frac{8\pi\left(\varepsilon_c+p_c\right)}{3}\left(W_c^{(2)}-C_{W2}^{\rm PF}\right), \nonumber\\
    & \lim_{\eta \to0} C_{V 2}^{\rm BDNK}=C_{V2}^{\rm PF}-\frac{5}{6}\left(W_c^{(2)}-C_{W2}^{\rm PF}\right),\\
     & \lim_{\eta \to0} C_{X 2}^{\rm BDNK}=C_{X2}^{\rm PF}+\frac{4\pi}{3}e^{\nu_c/2}\nonumber\\
     &\qquad\qquad\qquad\quad \times \left(3p_c^2+4 p_c \varepsilon_c +\varepsilon_c^2\right)\left(W_c^{(2)}-C_{W2}^{\rm PF}\right),\nonumber
\end{align}
where $C_{q 2}^{\rm PF}$ are the second-order Taylor coefficients of the perfect-fluid counterparts, defined though Eqs.~\eqref{eq:H1exp}--\eqref{eq:Vexp}. Together with Eq.~\eqref{eq:WBDNK}, these relations show that the second-order coefficients do not, in general, reduce to their perfect-fluid counterparts unless $W_c^{(2)} =C_{W2}^{\rm  PF}$. As in the odd-parity case, these conditions are satisfied within the small-transport-coefficient expansion by construction, which selects the branch connected to the perfect-fluid modes continuously.

Next, we consider the potential impact of the contributions neglected in the above analysis, such as the causal regulators. We assume thermal conductivity to be zero, i.e., $\kappa=0$. It is worth noting that, as shown in Eq.~\eqref{eq:NonAdiabaticCondition}, together with the expressions for the nonequilibrium amplitudes~\eqref{eq:S00BDNK}--\eqref{eq:S0ABDNK}, the causal regulators~$\tau_{\cal E}$,~$\tau_{\cal Q}$,~and $\tau_{\cal P}$ can, in general, induce entropy and/or composition perturbations. Hence, $\hat{\Sigma}$ and $\hat{\cal Y}^i$ are, in general, included in the perturbation variables. For simplicity, we assume that the background composition is homogeneous~$Y^i={\rm const.}$, and neglect its perturbation, i.e., $\Delta Y^i=0$, and hence ${\cal Y}^i = \hat{\cal Y}^i =0$. As discussed in Appendix~\ref{Appendix:BDNK}, a convenient set of variables for a practical implementation of the BDNK system~\eqref{eq:thermodynamicalEXSigma}--\eqref{eq:eqforSigmaYi} may be $(H,H_1,H_2,K,W,V,\hat{X},\hat{E},\Sigma)$ with the implicit relations for $\hat{W}$, $\hat{V}$, and $\hat{\Sigma}$ given in Eqs.~\eqref{eq:hatWhatV} and \eqref{eq:hatSigma}. Equation~\eqref{eq:eqforSigmaYi} can then be viewed as a second-order differential equation for $\Sigma$. We find that, when only $\tau_{\cal E}$ and $\tau_{\cal P}$ are present together with $\zeta$ and $\eta$, all the Taylor coefficients of the variables are characterized by $K_c$, $W_c$, and $W_c^{(2)}$ to sixth order in $r$, and they all reduce to the perfect-fluid counterparts in Eqs.~\eqref{eq:H1exp}--\eqref{eq:Vexp} in the limit~$\zeta,\eta,\tau_{\cal E},\tau_{\cal P}\to 0$, by setting $W_c^{(2)}=C_{W2}^{\rm PF}$. Thus, $\tau_{\cal E}$ and $\tau_{\cal P}$ give rise to no new unconstrained central amplitudes. By contrast, in the presence of $\tau_{\cal Q}$, we find one additional unconstrained parameter~$\Sigma_c^{(2)}$ defined through
\begin{align}
    \Sigma=\Sigma_c^{(2)}r^2+{\cal O}\left(r^4\right).
\end{align}
In the limit~$\zeta,\eta,\tau_{\cal E},\tau_{\cal P},\tau_{\cal Q}\to 0$, the second-order Taylor coefficients do not, in general, reduce to the perfect-fluid counterparts in Eqs.~\eqref{eq:H1exp}--\eqref{eq:Vexp}, unless $W_c^{(2)}=C_{W2}^{\rm PF}$ and $\Sigma_c^{(2)}=0$. We interpret that $\tau_{\cal Q}$ induces an entropy perturbation through Eq.~\eqref{eq:NonAdiabaticCondition}, which manifests itself as a new dynamical variable~$\Sigma$ through Eq.~\eqref{eq:Sigma}.

These observations suggest that the even-parity perturbations for BDNK viscous stars contain at least two additional unconstrained central parameters compared to the perfect-fluid case: $W_c^{(2)}$ induced by shear viscosity~$\eta$ and $\Sigma_c^{(2)}$ induced by the causal regulator~$\tau_Q$. These two additional unconstrained amplitudes suggest the requirements of two surface conditions, in addition to generalization of the normal-traction condition in the perfect-fluid case,~$\Delta p=0$. We expect that these additional conditions are the vanishing of the tangential traction and the normal energy flux associated with the out-of-equilibrium contribution at the surface. A traction vector is defined by $\mathbb{S}_{\mu\nu} n^\nu$, where $n^\mu$ is the unit normal to the surface and $\mathbb{S}_{\mu\nu}:=\Delta_\mu^\alpha \Delta_\nu^\beta (T_{\alpha \beta}+\mathscr{S}_{\alpha \beta})$, which explicitly reads
\begin{align}
    \mathbb{S}_{\mu\nu}=\left(p+{\cal P}\right)\Delta_{\mu \nu}+{\cal T}_{\mu\nu}.
\end{align}
The normal and tangential traction conditions are then
    \begin{align}
        \mathbb{S}_{\mu \nu} n^\mu n^\nu =0 &\Leftrightarrow p+{\cal P}+{\cal T}_{\mu \nu} n^\mu n^\nu=0,\label{eq:nomaltraction}\\
        \mathbb{S}_{A \mu} n^\mu =0 &\Leftrightarrow {\cal T}_{A \mu} n^\mu=0.\label{eq:tangentialtraction}  
    \end{align}
In the perfect-fluid case~$({\cal P}=0={\cal T}_{\mu \nu})$, the normal condition~\eqref{eq:nomaltraction} reduces to $\Delta p =0$ in the perturbed configuration, whereas the tangential condition~\eqref{eq:tangentialtraction} is satisfied identically. When bulk viscosity is the only out-of-equilibrium contribution, ${\cal T}_{\mu\nu}=0$, and therefore, it only contributes to the the normal condition~\eqref{eq:nomaltraction} through $\Delta p+ {\cal  P}=0$, without introducing a new independent surface condition. When shear viscosity is present, both the normal condition~\eqref{eq:nomaltraction} and the tangential condition~\eqref{eq:tangentialtraction} must be imposed at the surface. Furthermore, when $\tau_Q$ is present, the normal energy flux of the out-of-equilibrium correction,~${\cal Q}_\mu n^\mu$, must also vanish at the surface. This condition is equivalent to the vanishing of the nonequilibrium amplitude~$S_{01}$ at the surface.

To summarize, for the BDNK fluid with $\kappa=0$, the Taylor expansion of the perturbation variables around the center~$r=0$ is characterized by four unconstrained parameters~$K_c,W_c,W_c^{(2)},\Sigma_c^{(2)}$, where $W_c^{(2)}$ and $\Sigma_c^{(2)}$ are induced by shear viscosity and the causal regulator~$\tau_{\cal Q}$, respectively. With the outgoing-wave condition at infinity, if the above three surface conditions, i.e., vanishing of the normal and tangential traction and of the normal energy flux of the out-of-equilibrium contribution at the stellar surface, determine discrete values of $\omega$ and the ratios among $K_c,W_c,W_c^{(2)},\Sigma_c^{(2)}$ simultaneously, then new mode families may emerge. We leave a detailed analysis of this possibility for future work.

\section{Summary and discussion}\label{Section:Summary}
In this work, we have carried out a systematic, model-agnostic analysis of out-of-equilibrium effects in non-radial oscillations of spherically symmetric relativistic stars. To this end, we have constructed, to our knowledge, the first general framework for linear, non-radial relativistic stellar perturbations that incorporates generic nonequilibrium corrections to the perfect-fluid sector in both the even- and odd-parity channels, by extending the Lindblom-Detweiler formalism~\cite{Lindblom:1983ps,Detweiler:1985zz}. Our framework is formulated in terms of the tensorial structure and thermodynamic decomposition of generic corrections without relying on any specific constitutive relations. As an application, we have applied our formalism to BDNK fluids and analyzed the resulting shifts of the $f$- and $w$-modes.

In our formalism, out-of-equilibrium corrections are encoded in a set of radial functions, which we dub the {\it nonequilibrium amplitudes}. Leaving these amplitudes agnostic, the perturbation equations are presented in Eqs.~\eqref{eq:thermodynamicalEXSigma} --\eqref{eq:eqforSigmaYi} for the even-parity sector, and Eqs.~\eqref{eq:RWinterior} and~\eqref{eq:EqforUinpsi} for the odd-parity sector, together with the online supplemental material~\cite{OnlineLink}. Remarkably, the system in the even-parity sector retains a form close to the perfect-fluid case within the Lindblom-Detweiler formalism, with a modest number of additional terms. This is because the heat-flux contributions can be absorbed into redefined perturbation variables by exploiting the hydrodynamic frame ambiguity. It is worth emphasizing, however, that our procedure does not amount to a change of frame; rather, it merely rewrites the perturbation equations in a form in which the heat-flux sector does not appear explicitly. We also note that, for fluid models involving independent dynamical variables, such as M{\"u}ller-Israel-Stewart fluids~\cite{1967ZPhy..198..329M,Israel:1976tn,Israel:1976efz,1977RSPSA.357...59S,1979RSPSA.365...43I,Israel:1979wp}, the corresponding additional evolution or constraint equations must be supplied separately.

For the BDNK fluid, we first analytically showed that the causal-regulator contributions are structurally suppressed within a regular perturbative expansion in small transport coefficients around the perfect-fluid limit, and therefore enter only at higher order in the expansion. 
Within this perturbative scheme, we found that, for the $f$-mode, both shear and bulk viscous effects reduce the oscillation frequency and enhance the damping rate. The trend in the damping timescale can be interpreted as a consequence of viscosity providing an additional dissipative channel for the mode.
Moreover, the numerically computed shift in the damping rate is consistent with the independent estimates of Ref.~\cite{1987ApJ...314..234C} at the order-of-magnitude level, providing an independent check of our formalism, at least within the perturbative treatment based on the small-transport-coefficient expansion around the perfect-fluid limit. By contrast, for the $w$-modes, the damping rate is reduced by the viscous effects within the setup considered in this work.  

We found that viscous effects have a stronger impact on the damping rate than on the oscillation frequency. This trend has also been observed in previous studies of radial oscillations of BDNK stars~\cite{Shum:2025jnl,Keeble:2026bzo}. Another intriguing finding for a fixed stellar model and equation of state is that, within our particular parameterization of the radial profiles of transport coefficients and for the transport coefficients of the same order of magnitude, shear viscosity induces a larger shift than bulk viscosity by nearly two orders of magnitude in the $f$-mode case and by nearly three orders of magnitude in the $w$-mode case. The hierarchy between the shear- and bulk-viscous effects is notably different from the observation in radial stellar oscillations. In such cases, the expansion of fluid elements is expected to dominate over shear distortion, and hence, the bulk-viscous effects may play the leading role in mode shifts~\cite{Shum:2025jnl}. By contrast, in non-radial oscillations, shear distortion may be more efficiently induced because the perturbation is no longer spherically symmetric. This may provide a qualitative explanation for why shear viscosity has a stronger impact in our results. Note that, even for the same equilibrium star, the oscillation frequencies can differ depending on whether the matter composition remains frozen over the oscillation period or is always in $\beta$-equilibrium~\cite{Montefusco:2024xrx}. Therefore, the viscosity-induced frequency shifts reported here are relative to the perfect-fluid frequencies calculated under the assumptions used in this work.

The analysis of regular solutions at the stellar center shows that, in the even-parity sector, the BDNK contributions introduce at least two additional unconstrained central parameters in the Taylor expansion of the perturbation variables around the center, compared to the perfect-fluid case. This structure suggests the possible emergence of eigenfunctions with no regular perfect-fluid counterparts, and hence of new modes that do not reduce to perfect-fluid modes in the limit of vanishing transport coefficients. We emphasize that these conclusions do not establish the existence of new mode families, since our analysis is restricted to the local behavior of the regular perturbation variables around the stellar center, rather than solving the full eigenvalue problem. In the odd-parity sector, two unconstrained central amplitudes remain because the fluid-velocity perturbation becomes a dynamical variable. This suggests the emergence of a new family, consistent with the results of  Ref.~\cite{Bussieres:2026rnz}. It is also worth noting that the BDNK fluid model is constructed as a first-order hydrodynamic effective theory obtained by a gradient expansion about zeroth-order perfect-fluid constitutive relations. Therefore, these new branches should be interpreted as features of the closed eigenvalue problem, rather than robust controlled predictions within the effective-theory framework.

While the current study is restricted to static, nonrotating and cold background configurations, these assumptions are expected to be valid for NSs during most of the inspiral phase, where the stars are typically cold and slowly rotating. Consequently, the formalism developed here may already provide qualitative insight into dissipative corrections to oscillations and tidal responses in inspiraling binaries. In contrast, applications to post-merger remnants are less direct because such systems are hot, rapidly rotating, and strongly dynamical. Nevertheless, the formalism developed in this work provides a foundation for systematically extending relativistic perturbation theory beyond the perfect-fluid approximation and for investigating how dissipative processes may affect GW observables in more realistic astrophysical settings.

The present work opens several directions for future extensions. First, it would be useful to quantify the impact of thermal conductivity, which has been neglected in the present work, on mode shifts and modal stability. Second, a modal stability analysis beyond the present perturbative scheme adopted here would help further clarify the properties of the BDNK fluid, including search for a family of new modes especially in the even-parity sector that have not been discovered yet. Third, applying our formalism to other established fluid models, as well as theories developed in the future, would enable systematic comparisons of the mode structure among different models and may help assess the possible observational impact of out-of-equilibrium effects on stellar pulsations.


\begin{acknowledgments}
We are grateful to Vitor Cardoso, Arun Krishna Ganesan, Lorenzo Gavassino, Leonardo Gualtieri, Abhishek Hegade, Kostas Kokkotas, Jorge Noronha, Paolo Pani, and Jaime Redondo-Yuste for valuable discussions and useful correspondence.
T.K. is supported by the MUR FIS2 Advanced Grant ET-NOW (CUP:~B53C25001080001) and by the INFN TEONGRAV initiative.
K.Y. acknowledges support from NSF Grant PHY-2309066 and PHYS-2339969.

\end{acknowledgments}


\appendix
\section{Derivation of the even-parity perturbation equations}\label{Appendix:Derivation}
We provide detailed derivation steps of Eqs.~\eqref{eq:eqforH1}--\eqref{eq:eqforSigmaYi}. We obtain Eq.~\eqref{eq:eqforH1} by substituting Eq.~\eqref{eq:eqforH2} into $\delta G_{t \theta}=8\pi (\delta T_{t \theta}+\mathscr{S}_{t\theta})$. Equation~\eqref{eq:eqforK} arises from  $\delta G_{tr}=8\pi (\delta T_{tr}+\mathscr{S}_{tr})$. With these results, we directly obtain Eq.~\eqref{eq:eqforW} from $\delta (\nabla_\mu T^{\mu\theta})=-\nabla_\mu \mathscr{S}^{\mu \theta}$. To derive Eq.~\eqref{eq:eqforpX}, we first solve $\delta G_{r\theta}=8\pi (\delta T_{r\theta}+\mathscr{S}_{r\theta})$ to express $H_0'$ in terms of $(H,H_1,K,V,W)$, and then, substitute this expression into $\delta  (\nabla_\mu T^{\mu r})=-\nabla_\mu \mathscr{S}^{\mu r}$. Using these results, we then derive Eq.~\eqref{eq:eqforH} from $\delta G_{rr}=8 \pi (\delta T_{rr}+\mathscr{S}_{rr})$. Equation~\eqref{eq:eqforX} is obtained from Eq.~\eqref{eq:hatX} with Eq.~\eqref{eq:thermodynamicalrelation} by substituting the above results into all the terms involving the first-order derivatives. We finally obtain Eq.~\eqref{eq:eqforSigmaYi} from Eq.~\eqref{eq:hatSigma} with Eq.~\eqref{eq:NonAdiabaticCondition}. Thus, the derivation of the set of the perturbation equations equivalent to Eqs.~\eqref{eq:fieldeqs} and~\eqref{eq:baryonnumberconservation} is complete.

\section{Nonequilibrium amplitudes for the BDNK fluid}\label{Appendix:BDNK}

Here, we provide the expressions for the nonequilibrium amplitudes within the BDNK fluid. The same expression can be found online~\cite{OnlineLink}.

\subsection{Even-parity sector}
For the even-parity sector, the nonequilibrium amplitudes are given by 
\begin{widetext}
\allowdisplaybreaks
    \begin{align}
S_{00}=&-i\omega \tau_{\varepsilon}\frac{e^{-\nu/2}}{r^{\ell-2}}  \Xi,\label{eq:S00BDNK}\\
S_{01}=&-\kappa\frac{\varepsilon+p}{n}\frac{e^{\nu/2}}{r^{\ell-2}}\left[\frac{\mu_i T'}{T^2}\delta T_r-\left(\frac{T'}{T}\delta \mu_{r,i}+\frac{\mu_i}{T}\delta T_r'\right)+\delta \mu_{r,i}'\right]\nonumber\\
&-\tau_{\cal Q}\frac{e^{\nu/2}}{r^{\ell-2}} 
\Bigg[e^\lambda R_r+\frac{\Xi_\alpha }{2r} 
\left\{c_s^2 \left(1-e^{\lambda} \left(1+8\pi r^2 p\right) \right)
-2r\left(c_s^2\right)'\right\} \label{eq:S01BDNK}\\
&-\frac{\Xi}{2r}\left\{1+c_s^2-e^{\lambda}\left(1+c_s^2\right)\left(1+8\pi r^2 p\right)-2r\,\left(c_s^2\right)'\right\}
+ c_s^2 \left(\Xi'-\,\Xi'_\alpha\right)
\Bigg],\nonumber\\
S_{0A}=&-\kappa \frac{e^{\nu/2}}{r^{\ell-1}}\frac{\varepsilon+p}{n}\left(\delta \mu_{r,i}-\mu_i \frac{\delta T_r}{T}\right) -\tau_{\cal Q} \frac{e^{\nu/2}}{r^{\ell-3}}\left[R_A+\frac{c_s^2}{r^2}\left(\Xi-\Xi_\alpha\right)\right],\label{eq:S0ABDNK}\\
S_{0}=&-i \omega \tau_{\cal P} \frac{e^{-\nu/2}}{r^{\ell-2}} \Xi\nonumber\\
&
-\frac{i\omega \eta}{3}e^{-(\lambda+\nu)/2}\left[2e^{\lambda/2}r^2 H_2-2 e^{\lambda/2} r^2 K+ 2 e^{\lambda/2}\ell\left(\ell+1\right)V-4\left(\ell-2\right)W-4r W'\right]
\label{eq:S0BDNK}\\
&
-\frac{i \omega \zeta}{2}e^{-(\lambda+\nu)/2}\left[e^{\lambda/2}r^2 H_2+2 e^{\lambda/2} r^2 K-2 e^{\lambda/2}\ell\left(\ell+1\right)V-2\left(\ell+1\right)W-2r W'\right]
,\nonumber\\
S_{1}=&-i\omega \eta e^{-\nu/2}\left[\left(\ell-2\right)V+r V'-e^{\lambda/2}W\right],\\
S_{Z}=&-2i \omega \eta e^{-\nu/2} V,\\
S_{\Omega}=&-i \omega \tau_{\cal P} \frac{e^{-\nu/2}}{r^{\ell-2}} \Xi\nonumber\\
&
-\frac{i\omega \eta}{3}e^{-(\lambda+\nu)/2}\left[-e^{\lambda/2}r^2 H_2+ e^{\lambda/2} r^2 K- e^{\lambda/2}\ell\left(\ell+1\right)V+2\left(\ell-2\right)W+2r W'\right]\label{eq:SOmegaBDNK}\\
&
-\frac{i \omega \zeta}{2}e^{-(\lambda+\nu)/2}\left[e^{\lambda/2}r^2 H_2+2 e^{\lambda/2} r^2 K-2 e^{\lambda/2}\ell\left(\ell+1\right)V-2\left(\ell+1\right)W-2r W'\right],\nonumber\\
J_0=&0,\\
J_1=&0,\\
J_A=&0.
\end{align}
\end{widetext}
Here, we have imposed $\mu_i/T={\rm const.}$, requiring that the heat flux vanishes in equilibrium. Moreover, $\Xi_\alpha=\Xi_\alpha(r)$ is defined in Eq.~\eqref{eq:Xialpha}, and reads
\begin{align}
    \Xi_\alpha=-\alpha_1 n T e^{-\nu/2} r^\ell \Sigma- n \mu_i \alpha_{2i} e^{-\nu/2} r^\ell {\cal Y}^i.
\end{align}
We also define radial functions~$\delta T_r$ and $\delta \mu_{r,i}$ by $\delta T=\delta T_r e^{-i \omega t} Y_{\ell m}$ and $\delta \mu_i=\delta \mu_{r,i} e^{-i \omega t} Y_{\ell m}$, respectively. The radial functions $R_r$ and $R_A$ are defined through
\begin{align}
    {\cal R}^\mu = \left(0,R_r Y_{\ell m}, R_A E_\theta^{\ell m},  R_A \frac{E_\varphi^{\ell m}}{\sin^2\theta}\right)e^{-i \omega t},\label{eq:vectorR}
\end{align}
which is defined by Eq.~\eqref{eq:Rmu}, and
\begin{widetext}
 \begin{align}
    R_r=&\frac{\varepsilon+p}{2}r^{\ell-1} e^{-\lambda} \Bigg[
\ell H+rH'
+2e^{-\nu}r^2\omega^2 H_1
+\left(\frac{1}{2}H_2+K\right)
\left(
e^{\lambda}-1+2\ell c_s^2+2r\left(c_s^2\right)'
+8\pi e^{\lambda} r^2 p
+\frac{2r c_s^2}{\varepsilon+p}\varepsilon'
\right)
\nonumber\\
&-\frac{\ell(\ell+1)}{r^2}V
\left\{
e^{\lambda}-1+2(\ell-2)c_s^2+2r\left(c_s^2\right)'
+8\pi  e^{\lambda}r^2 p
+\frac{2r c_s^2}{\varepsilon+p}\varepsilon'
\right\}
\nonumber\\
&-\frac{e^{-\lambda/2-\nu}}{2r^2}W
\Bigg\{
e^{2\lambda+\nu}\left(1+8\pi r^2 p\right)^2
+2e^{\lambda}\left(
2r^2\omega^2
+e^{\nu}\left(
3+(\ell+1) c_s^2
-8\pi r^2\left(1+(\ell+1)c_s^2\right)\varepsilon
\right)
\right)\label{eq:Rr}\\
&-e^{\nu}\left(
7-2(\ell+1)(2\ell-5)c_s^2
-4(\ell+1)r\left(c_s^2\right)'
\right)
+\frac{4e^{\nu}(\ell+1)r c_s^2}{\varepsilon+p}\,\varepsilon'
\Bigg\}
\nonumber\\
&+\frac{e^{-\lambda/2}}{r}W'
\left\{
c_s^2\left(
1-4\ell+e^{\lambda}\left(-1+8\pi r^2\varepsilon\right)
\right)
-2r\left(c_s^2\right)'
-\frac{2r c_s^2}{\varepsilon+p}\,\varepsilon'
\right\}
\nonumber\\
&\left.+2r c_s^2\left\{\frac{1}{2}H_2'+K'-\frac{\ell(\ell+1)}{r^2}V'-\frac{1}{r}e^{-\lambda/2}W''
\right\}\right],\nonumber
\end{align}
and
\begin{align}
    R_A=&\frac{\varepsilon+p}{2}r^{\ell-2}\left[H+c_s^2 H_2+2 c_s^2K+\frac{2e^{-\nu}}{r^2}\left\{r^2 \omega^2-\ell \left(\ell+1\right)e^\nu c_s^2\right\}V\right.\nonumber\\
        &\left.+\frac{e^{-\lambda/2}}{r^2}\left\{e^\lambda-1-2\left(\ell+1\right)c_s^2+8\pi e^\lambda  r^2 p\right\}W-\frac{2e^{-\lambda/2}}{r}c_s^2W'\right].
\end{align}
\end{widetext}

Let us now briefly outline a practical procedure for solving the BDNK perturbation system. For simplicity, we assume that the background composition is homogeneous, i.e., $Y^i={\rm const.}$, and its perturbation is absent, i.e., $\Delta Y^i=0$, and hence ${\cal Y}^i=\hat{\cal Y}^i=0$. Using the definition of $\Xi$ in Eq.~\eqref{eq:TDeltasmuDeltaY}, together with the first law of thermodynamics~\eqref{eq:firstlaw}, we obtain
\begin{align}
    \Xi= - nT e^{-\nu/2}r^\ell \Sigma,\label{eq:Xi}
\end{align}
and hence $\Xi_\alpha=\alpha_1 \Xi$. These relations allow one to express the contributions associated with $\tau_{\cal E}$, $\tau_{\cal P}$, and $\tau_Q$ in terms of $\Sigma$. Then, Eq.~\eqref{eq:eqforSigmaYi} arising from the definition of $\Sigma$ given in Eq.~\eqref{eq:Sigma} with Eq.~\eqref{eq:NonAdiabaticCondition} can be viewed as a second-order differential equation for $\Sigma$. Once $S_{01}$ is determined by $\Sigma$, the redefined variable~$\hat{\Sigma}$ is reconstructed by Eq.~\eqref{eq:hatSigma}. Thus, a convenient set of nine variables is~$(H,H_1,H_2,K,W,V,\hat{X},\hat{E},\Sigma)$, while $\hat{V}$, $\hat{W}$, and $\hat{\Sigma}$  are related to these variables through the implicit relations in Eqs.~\eqref{eq:hatWhatV} and~\eqref{eq:hatSigma}. The nine equations, Eqs.~\eqref{eq:thermodynamicalEXSigma}--\eqref{eq:eqforSigmaYi}, are then solved simultaneously together with these implicit relations.

\subsection{Odd-parity sector}
For the odd-parity sector, the nonequilibrium amplitudes are given by
\begin{widetext}
\begin{align}
    S_{0A}^-=&-\tau_{\cal Q} e^{\nu/2} r^2 R_A^- ,\label{eq:BDNKS0AOdd}\\
      S_{1A}^-=&\eta e^{-\nu/2}\left[i\omega h_1-\frac{e^\nu}{4\pi\left(\varepsilon+p\right)}\left\{U'+\left(\nu'-\frac{2}{r}-\frac{\varepsilon'+p'}{\varepsilon+p}\right)U\right\}\right],\label{eq:BDNKS1AOdd}\\
        S_{Z}^-=&-\eta \frac{e^{\nu/2}}{2\pi \left(\varepsilon+p\right)}U.\label{eq:BDNKSZOdd}
\end{align}
\end{widetext}
with 
\begin{align}
   {\cal R}^\mu=&R_A^-\left(0,0,B_\theta^{\ell m},\frac{B_\varphi^{\ell m}}{\sin^2\theta}\right)e^{-i \omega t},\label{eq:vectorROdd}\\
   R_A^-=&-\frac{i \omega}{4\pi r^2} \left[4\pi e^{-\nu}(\varepsilon+p) h +U \right].\nonumber
\end{align}

\bibliography{Ref}

\end{document}